\documentclass{aa} 
\usepackage{graphicx}
\usepackage{txfonts}
\usepackage{longtable,lscape}
\usepackage{natbib}
\bibpunct{(}{)}{;}{a}{}{,}
 
\begin{document}
%
   \title{FLAMES spectroscopy of low-mass stars in the 
   young clusters $\sigma$~Ori and $\lambda$~Ori\thanks{Based on Data
   collected at the ESO Very Large Telescope, Paranal Observatory, Chile 
   [programs 074.D-0136(A) and 076.C-0125(A)]}}

   \subtitle{}

   \author{G. G. Sacco 
          \inst{1}\fnmsep\inst{2}
          \and
         E. Franciosini\inst{1}
	   \and
 S. Randich\inst{4}
 \and
	  R. Pallavicini\inst{1}\fnmsep\inst{3}
	  }

   \offprints{G.G. Sacco, \email{sacco@astropa.inaf.it}}

   \institute{INAF, Osservatorio Astronomico di Palermo, Piazza del
   Parlamento, 1, 90134 Palermo, Italy\\
\and Consorzio COMETA, Via S. Sofia, 64, 95123 Catania, Italy \\ 
e-mail: sacco@astropa.inaf.it\\
\and INAF, Headquarters, Viale del Parco Mellini 84, 00136 Roma, Italy\\
   \and INAF, Osservatorio Astrofisico di Arcetri, Largo E. Fermi, 50125
	   Firenze, Italy }

   \date{Received 12 November 2007/ accepted .....}

 
  \abstract
  {}
  {We performed a detailed membership selection and studied the accretion
properties of low-mass stars in the two apparently very similar young
(1-10 Myr) clusters $\sigma$~Ori and $\lambda$~Ori. }
  {We observed 98 and 49 low-mass ($0.2-1.0 M_{\sun}$) stars in
$\sigma$~Ori and $\lambda$~Ori respectively, using the multi-object optical
spectrograph FLAMES at the VLT, with the high-resolution ($R\sim$17,000)
HR15N grating (6470$-$6790~\AA). We used radial velocities, Li and
H$\alpha$ to establish cluster membership and H$\alpha$ and other optical
emission lines to analyze the accretion properties of members.}
  {We identified 65 and 45 members of the $\sigma$~Ori and $\lambda$~Ori
clusters, respectively and discovered 16 new candidate binary
systems. We also measured rotational broadening for 20 stars
and estimated the mass accretion rates in 25 stars of
the $\sigma$~Ori cluster, finding values between $10^{-11}$ and 
$10^{-7.7} M_{\odot}$~yr$^{-1}$ and in 4 stars of the $\lambda$~Ori cluster,
finding
values between $10^{-11}$ and $10^{-10.1} M_{\odot}$~yr$^{-1}$. Comparing
our results with the infrared photometry obtained by the \textit{Spitzer}
satellite, we find that the fraction of stars with disks and the
fraction of active disks is larger in the $\sigma$~Ori cluster (52$\pm$9\%
and 78$\pm$16\%) than in $\lambda$~Ori (28$\pm$8\% and 40$\pm$20\%).
  }
  {The different disk and accretion properties of the two clusters could be 
 due either to the effect of the high-mass stars and the supernova explosion 
in the $\lambda$~Ori cluster or to different ages of the cluster
populations. Further observations are required to draw a 
definitive conclusion. }

   \titlerunning{FLAMES spectroscopy in the $\sigma$~Ori and $\lambda$~Ori
clusters }
   \maketitle
%

\section{Introduction}

The time evolution of accretion properties and disk dissipation mechanisms
in young stars are two of the main open issues in the star formation theory.
The analysis of the accretion properties of the stellar populations
belonging to young clusters, using high-resolution ($R\sim$20,000)
spectroscopy, is a powerful tool to investigate these questions. Among young
clusters, $\sigma$~Ori and $\lambda$~Ori are two of the richest clusters
near the Sun in the age range (1$-$10 Myr) during which young stars
lose their circumstellar disk and stop accreting material from it
\citep{Hartmann1998ApJ,Haisch2001ApJ,Sicilia2005AJ, Sicilia2006ApJ}.     

The $\sigma$~Ori cluster was discovered by the \textit{ROSAT} satellite
\citep{Wolk1996PhDT, Walter1997MmSAI} around the O9.5~V binary star
$\sigma$~Orionis AB (distance $352^{+166}_{-85}$~pc,
\citealt{Perryman1997A&A}). Because of its proximity and very low reddening
\citep{Oliveira2004MNRAS}, during the last decade $\sigma$~Ori has become
one of the best studied young clusters. Its low-mass and substellar
population, extending down to planetary-mass objects, has been extensively
observed by photometry and low-resolution spectroscopy, both in infrared and
optical bands \citep{Zapatero2000Sci,Bejar2001ApJ,Zapatero2002A&A,
Barrado2003A&A, Scholz2004A&A, Sherry2004AJ, Bejar2004AN,
Burningham2005MNRAS,Kenyon2005MNRAS, Caballero2007A&Ab}, while its high-mass
stellar content has been studied by \cite{Caballero2007A&A}. The estimated
median age of the cluster ranges from 1 to 8 Myr, depending on the assumed
distance and measurement method adopted by different authors. Using
measurements of tangential and radial velocities, \cite{Jeffries2006MNRAS}
and \cite{Caballero2007A&A} argued that, in the same region of the
$\sigma$~Ori cluster, there is a sparser, kinematically separate young
stellar population belonging to the Orion OB1b association.
\cite{Zapatero2002A&A}, using low-resolution spectroscopy of a sample of 27
low-mass and substellar objects, estimated a fraction of about 30$-$40\% of
classical T Tauri stars (CTTSs), while the fraction of circumstellar disks
(33$\pm$6\%) was first determined by
\cite{Oliveira2004MNRAS,Oliveira2006MNRAS}, using infrared photometric data
in the $K$ and $L$ bands. Recently, \cite{Hernandez2007ApJ} used a near-infrared
survey carried out by the \textit{Spitzer} satellite to investigate the disk
properties of 336 candidate members, finding that 34\% of them harbor a
circumstellar disk (27\% having thick disks and 7\% having evolved optically
thin disks). Moreover, they found that the total fraction of stars with
disks decreases with increasing mass from 36\% (31\% with thick disks) for
low-mass T Tauri stars to 15\% (4\% with thick disks) for Herbig Ae/Be
stars. Finally, the X-ray properties of high-mass and low-mass cluster
members  have been studied by \cite{Sanz-Forcada2004A&A} and
\cite{Franciosini2006A&A} using the \textit{XMM-Newton} satellite. 

$\lambda$~Orionis is an O8~III star (distance 400~pc,
\citealt{Murdin1977MNRAS}) which excites the HII region S264, delimited by a
dust ring with a diameter of 9$^{\circ}$ discovered by the \textit{IRAS}
satellite \citep{Zhang1989A&A}. The cluster, distributed over an area  of 1
square degree around the O8 star, was discovered by \cite{Gomez1998AJ}
together with two clusters located in the nearby clouds B30 and B35, by
analyzing the spatial correlation among the H$\alpha$ sources discovered by
\cite{Duerr1982ApJ}. \cite{Dolan1999AJ}, using medium-resolution
spectroscopy, selected 72 members brighter than $R=16$ around
$\lambda$~Orionis and  found that the fraction of CTTSs (7\%) belonging to
the cluster is very low if compared to other clusters and star-forming
regions in the same age range (1$-$10~Myr). \cite{Dolan2001AJ, Dolan2002AJ}
extended this analysis to the whole region and suggested that the star
formation process started 8$-$10~Myr ago, with an accelerating star-formation
rate, and stopped 1$-$2~Myr ago after a supernova explosion, which shredded
the central cloud and formed the current gas ring. They also suggested that,
before the supernova explosion, the cluster was still bound to its natal
cloud, therefore the young stars closer    to the central OB stars,
including also the supernova progenitor, lost their circumstellar disks due
to photoevaporation by the far-ultraviolet emission from the high mass
stars. \cite{Barrado2004ApJ} selected 170 candidate members through deep
optical photometry and performed low-resolution spectroscopy of 33 very
low-mass and substellar objects. The properties of circumstellar disks have
been studied by \cite{Barrado2007ApJ} by means of the \textit{Spitzer}
satellite. They found that the total fraction of members with disks (both
thick and evolved optically thin disks) is 31\%, but the fraction of stars
with a thick disk is only 14\%.   They also found that the distribution of
Class II stars is inhomogeneous, namely, most of them are located in a
filament that goes from $\lambda$~Orionis to the B35 cloud. Moreover, since
several Class II stars are located near the cluster center, they argued that
high-mass stars and the supernova explosion had no effect on the
circumstellar disks.     

Although the two clusters have been extensively observed during the last
decade, only few spectroscopic data at a resolution equal to or higher than
$R=10,000-15,000$  are available and, especially for the $\sigma$~Ori cluster,
most of the cluster members have been selected by means of photometry only.
Therefore, the catalogues of members of both clusters are likely
contaminated by foreground field stars or young stellar objects belonging to
different populations, and the accretion properties of the known members are
poorly determined.

We performed high-resolution spectroscopy of two large samples of stars in
the $\sigma$~Ori and $\lambda$~Ori clusters, using FLAMES at the VLT
\citep{Pasquini2002Messanger}, in order to select new high-probability
cluster members, to identify CTTSs using H$\alpha$ emission  and other
accretion indicators and to measure mass accretion rates (MARs). Moreover,
the comparison between the ``twin'' clusters $\sigma$~Ori and $\lambda$~Ori is
a powerful tool to investigate the origin of the presumed lack of CTTSs in
the $\lambda$~Ori cluster.  In \cite{Sacco2007A&A} we reported a first
important result for the $\sigma$~Ori cluster, obtained from the analysis of
these data, namely, the discovery of three Li-depleted stars, with
isochronal and nuclear ages greater than 10$-$15~Myr. In this paper, we focus
on the other results obtained by these data.

The paper is organized as follow: in Sect.~\ref{par:obs} we describe target
selection,
observations and data analysis; results are presented in
Sect.~\ref{par:results} and
discussed in Sect.~\ref{par:discussion}, where we also perform a comparison
with the infrared
data obtained by the \textit{Spitzer} satellite. Conclusions are summarized
in Sect.~\ref{par:conclusions}.

\section{Observations and data analysis}
\label{par:obs}

\subsection{Target selection \label{par:T_selection}}

\begin{figure*}
\centering
\includegraphics[width=17 cm]{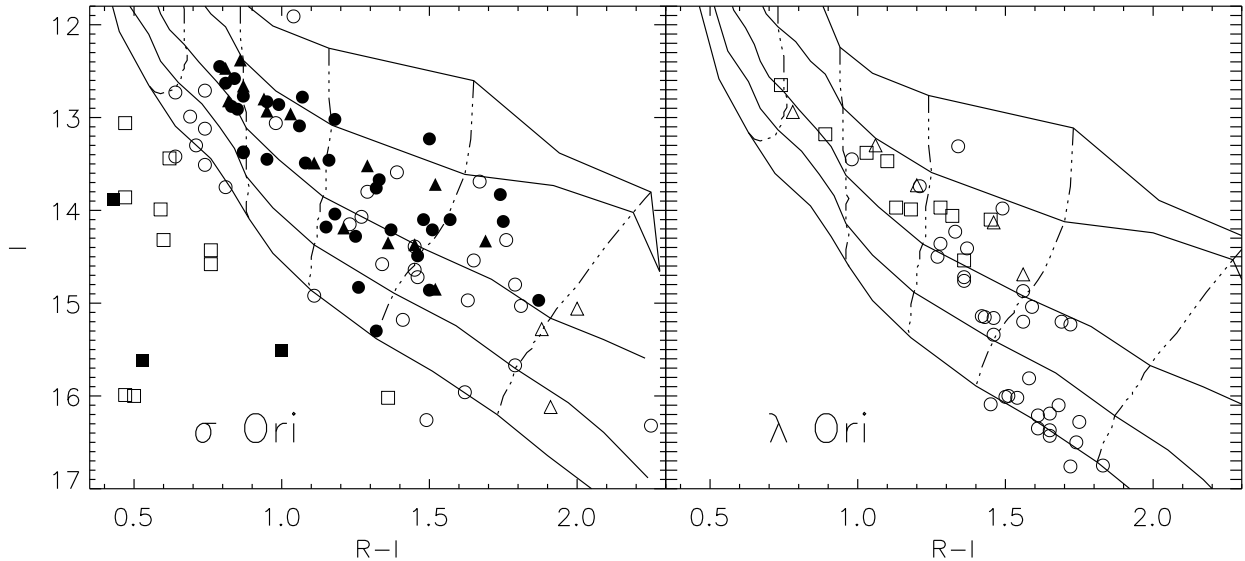} 
\caption{Color-magnitude diagrams of the selected objects in the
$\sigma$~Ori (left panel) and 
$\lambda$~Ori (right panel) clusters. In the left panel, triangles
represent the probable members on the basis of low-resolution spectroscopy,
circles represent the candidate members on the basis of photometry and squares
represent the non-members on the basis of photometry; filled symbols mark 
sources with an X-ray counterpart. In the right panel, triangles 
represent the sources included only in the \cite{Dolan1999AJ} catalogue,
circles represent the sources included only in the  \cite{Barrado2004ApJ}
catalogue
and squares represent the sources included in both catalogues. 
Continuous and dotted lines are, respectively, isochrones at 1, 2, 5, 10 and
20 Myr and  tracks at 0.2, 0.3, 0.4, 0.6, 1.0 $M_{\sun}$ from the
\cite{Siess2000A&A} models, using the \cite{Kenyon1995ApJS} table for 
the conversion of colors and magnitudes into temperatures and luminosities.
Isochrones and tracks are corrected for the distance modulus 
(7.72 and 8.01 for $\sigma$~Ori and $\lambda$~Ori, respectively) and 
in the case of $\lambda$~Ori also for reddening, considering
a color excess $E(B-V)=0.12$ \citep{Diplas1994ApJS} and using the 
\cite{Munari1996A&A} transformation laws; for the $\sigma$~Ori cluster, 
as discussed by  \cite{Oliveira2004MNRAS},  reddening is
negligible.}
\label{fig:targetCMD}
\end{figure*}

We selected sources in the $\sigma$~Ori and $\lambda$~Ori 
cluster regions in the spectral range K6$-$M5, which, for a 5~Myr cluster 
at a distance of about 400~pc, corresponds to the magnitude ranges  13$<R<$18 
and 11$<J<$15.
All selected sources, except 4 objects without optical photometry, are
plotted in Fig.~\ref{fig:targetCMD},
where different symbols indicate the membership information available 
before this work.

For the $\sigma$~Ori cluster, targets were selected using 
optical, infrared and X-ray data. Optical and infrared data 
were retrieved from the literature \citep{Wolk1996PhDT, Zapatero2002A&A, 
Sherry2004AJ, Kenyon2005MNRAS} and the 2MASS 
(2 Micron All Sky Survey) catalogue 
\citep{Skrutskie2006AJ}. X-ray data were retrieved from 
\cite{Franciosini2006A&A}. 
We selected a total of 98 objects; of these, 18 are probable members on the
basis of low-resolution spectroscopy, 66 are candidate members 
on the basis of optical and/or infrared photometry compatible with an age
$\la$10~Myr 
(taking also into account the error on distance) and the remaining 14 objects
appear to be non-members based on photometry, but have no spectroscopic 
information. Of the 98 targets, 53 were also detected in X-rays.

In the $\lambda$~Ori cluster we observed 49 sources, included in 
the catalogues by \cite{Dolan1999AJ} (5 objects), by \cite{Barrado2004ApJ}
(34 objects) or by both (10 objects). The stars included in the
\cite{Dolan1999AJ} 
catalogue are probable members on the basis of 
medium-resolution spectroscopy, while the other stars  are candidate members
based on photometric selection, with 3 objects confirmed as members also by 
low-resolution spectroscopy.   

The sources selected in the $\sigma$~Ori and $\lambda$~Ori clusters are
listed in Table~\ref{tab:mag_star} and in Table~\ref{tab:star_lambda},
respectively. Optical and infrared magnitudes are retrieved from the
literature; spectral types for 18 stars are from previous spectroscopic
studies \citep{Zapatero2002A&A, Barrado2004ApJ}, while for the other stars
they have been derived from the $R-I$ color, using the scale of
\cite{Kenyon1995ApJS} interpolated for each half subtype. We did
not derive spectral types from our data,
because the spectral range covered by our spectra does not include enough
spectral features for the classification, and because veiling affecting
accreting objects prevents the use of spectroscopic indices based on flux
ratios. For the stars with a spectroscopic classification already performed
by other authors, we checked that the discrepancy between photometric and
spectroscopic spectral type is 
$\leq$1 subtype for 12 stars, between 1 and 2 subtypes for 4 stars 
and of 2.5 subtypes in 2 cases (S07 and L44).

We cross-correlated our list of targets with the catalogues of stars
observed by the \textit{Spitzer} satellite, published by
\cite{Hernandez2007ApJ} for $\sigma$~Ori and by \cite{Barrado2007ApJ} for
$\lambda$~Ori.
We have 83 sources in common with the \cite{Hernandez2007ApJ} catalogues, 78
with the catalogue of probable members and 5 with the catalogue of uncertain
members, 
and 44 sources in common with the \cite{Barrado2007ApJ} catalogue.
Taking into account the classification based on the slope $\alpha$ of the
spectral energy distributions in the 3.6$-$8.0~$\mu$m spectral range, in the
$\sigma$~Ori cluster sample there are 34 stars harboring a circumstellar
disk ($\alpha > -2.56$) and 49 diskless stars, while in the $\lambda$~Ori
sample there are 11 stars with a circumstellar disk and 33 diskless stars.

\subsection{Observations and data reduction}

Observations were carried out using the fiber-fed multi-object spectrograph 
FLAMES (Fiber Large Array Multi Element Spectrograph), mounted on the UT2
telescope at the VLT \citep{Pasquini2002Messanger} and operated in the
MEDUSA mode (132 fibers, each with an aperture of 1.2~arcsec on the
sky). We used for both clusters the high resolution HR15N grating
(6470$-$6790~\AA, spectral resolution $R=$17,000 and nominal dispersion
0.1~\AA~pixel$^{-1}$), which includes the lithium line at 6708~\AA, the
H$\alpha$ line (6563~\AA) and other emission lines indicative of accretion
and outflow phenomena (NII at 6583~\AA, HeI at 6678~\AA and SII at 6716
and 6731~\AA). For $\lambda$~Ori only, we used also the HR21 grating
(8480$-$9000~\AA, spectral resolution $R=$16,200 and nominal dispersion
0.13~\AA~pixel$^{-1}$), which includes the Ca II infrared triplet, but,
since no star in our sample shows signs of Ca II emission due to
accretion phenomena (see Sect.~\ref{par:acc}), these spectra have been used
only for the measurement of the radial velocity (RV). For both clusters the
FLAMES field of view (diameter 25\arcmin) was centered around the high-mass
central star which gives the name 
to the cluster (RA=5h 38m 48.9s, DEC= -2d 34m 22s  
and RA=5h 35m 06.5s, DEC= 9d 54m 0.0s, Equinox J2000, 
for $\sigma$~Ori and $\lambda$~Ori, respectively).

Observations were performed in service mode and were 
divided into separate runs of 1 hour duration each, including 
instrument overheads. The $\sigma$~Ori cluster was observed 
in 6 runs in October and December 2004, 
while $\lambda$~Ori was observed in 8 runs in
October and November 2005.  
The observation log is reported in Table~\ref{tab:OB}.

\setcounter{table}{2}
\begin{table}[hbt]
\caption{\label{tab:OB} Observation log.}
\centering
\begin{tabular}{cccc}
\hline
\hline
cluster&  setup  &  exposure & run dates  \\
 &  &  time$^{\,a}$ 
(m) &  \\

\hline
$\sigma$ Ori &  HR15N &43.2 &
2004-10-01 \\
 $\sigma$ Ori & HR15N & 40.3 &  
 2004-12-01 \\
 $\sigma$ Ori & HR15N & 43.2 &  
 2004-12-02 \\
 $\sigma$ Ori &  HR15N & 43.2 &  
 2004-12-03 \\
 $\sigma$ Ori & HR15N & 43.2 &  
 2004-12-04 \\
 $\sigma$ Ori & HR15N & 43.2 & 
  2004-12-09 \\
\hline
$\lambda$ Ori & HR15N & 46.0 &
2005-10-15 \\
$\lambda$ Ori & HR15N & 46.0  &  
2005-10-16 \\
$\lambda$ Ori & HR15N &46.0  &  
2005-10-17 \\
$\lambda$ Ori &  HR15N & 46.0  &  
2005-11-10 \\
$\lambda$ Ori &  HR21 &46.0 &
2005-11-12 \\
$\lambda$ Ori & HR21 &46.0  &  
2005-11-12 \\
$\lambda$ Ori & HR21 & 46.0  &  
2005-11-12 \\
$\lambda$ Ori &  HR21 & 46.0  &  
2005-11-23 \\

\hline
\multicolumn{4}{l}{$a$: Excluding instrument overheads}
\end{tabular}
\end{table}

\begin{figure*}
\centering
\includegraphics[width=17 cm,clip]{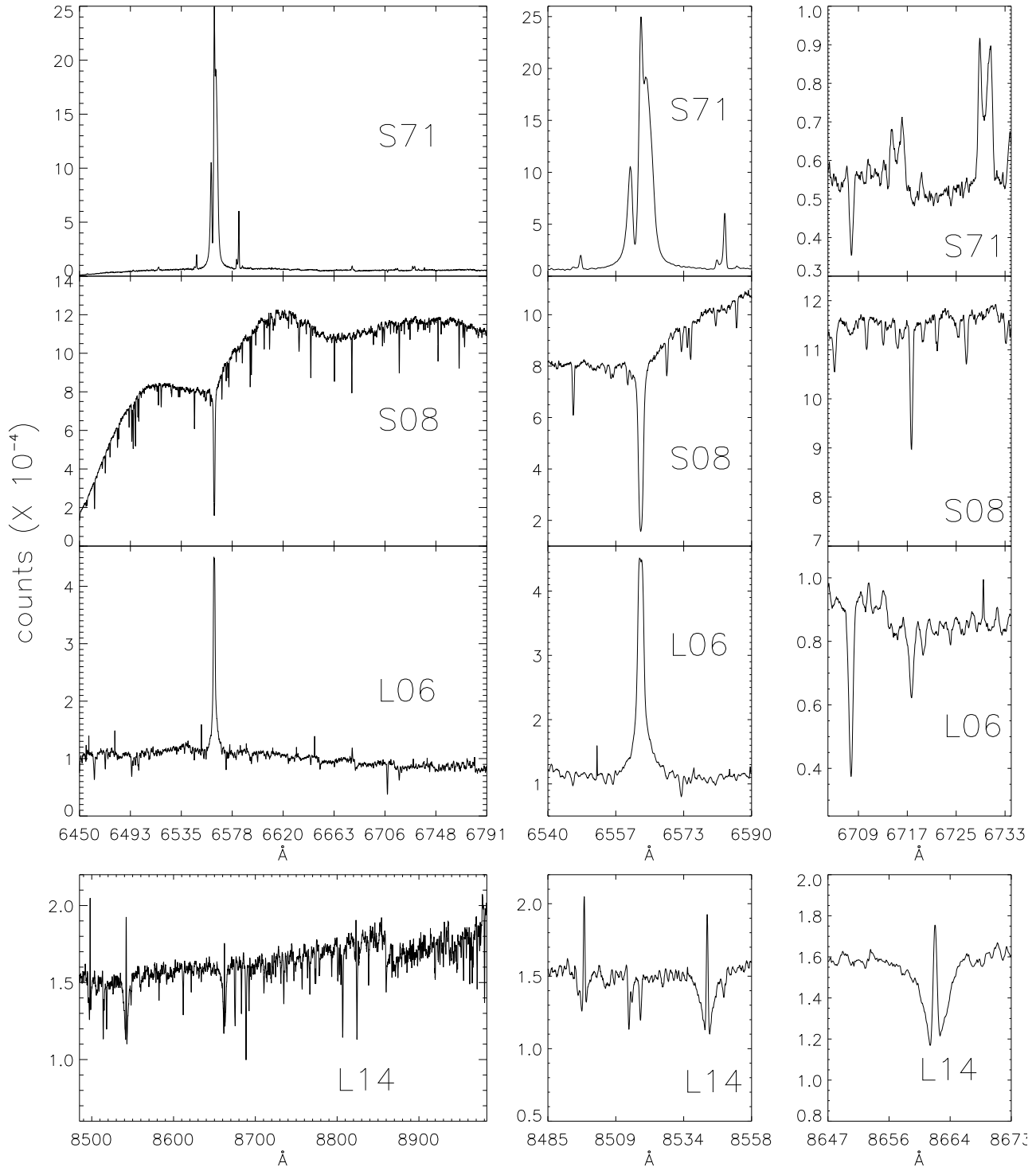}
\caption{Spectra of 4 stars of the sample. 
The three upper rows show spectra obtained with the HR15 grating:
in the left panels we show the whole spectral range, while the middle and
right panels show the range around the H$\alpha$ and the Li line,
respectively. 
For star S71, in the right panel the SII doublet at 6716 and 6731~\AA\ 
can also be seen.
The bottom row show a spectrum obtained with the HR21 grating: the whole
spectrum is shown in the left panel, while the middle and right panels
show the range around the CaII triplet. 
}
\label{fig:fig_sp_red}
\end{figure*}

Data reduction was performed using the GIRAFFE girBLDRS 
pipeline vers. 1.12, following the standard steps \citep{Blecha2004Man},
which include bias subtraction, division by a normalized flat-field,
correction for the differences in the fiber transmission, and wavelength calibration 
using a dispersion solution from a Thorium-Argon arc lamp. The spectra
processed by the pipeline are not corrected for instrumental response and sky
background, 
but, except for the RVs, all measurements were performed on the sum of the spectra recorded in the different runs, after subtracting a sky background spectrum.
The sky background was calculated independently 
for each observing run, using a set of fibers homogeneously 
distributed over the field (10 for $\sigma$~Ori and 20 for $\lambda$~Ori).  
For each run, we averaged a subsample of sky spectra without 
intense spikes. To increase the signal-to-noise ratio (S/N) for line
measurements, we then summed together, for each source, the spectra from the
different runs.
The S/N of the summed spectra ranges from $\sim$10 for stars with
$I\sim$16, to
$\sim$90 for stars with $I\sim$14 and to $\sim$170 for stars with $I\sim$12.
Some examples of stellar spectra are shown in Fig.~\ref{fig:fig_sp_red}.

\subsection{Radial velocities}

We measured, independently, the RVs in each observing 
run, with the aim of identifying cluster members and selecting 
candidate binary systems.
RVs were measured by Fourier cross-correlation \citep{Tonry1979AJ}, using 
the IRAF\footnote{IRAF is 
distributed by the National Optical Astronomy Observatory, which is operated by
the Association of Universities for Research in Astronomy, Inc.,under contract to
the National Science Foundation.} task FXCOR.
In brief, all the spectra were cross-correlated with one template spectrum 
chosen among the stars belonging to the observed sample (stars S07 and S56 for $\sigma$~Ori and
L04 and L17 for $\lambda$~Ori). Specifically, stars S07 and L04 have been used only for earlier type stars,
and stars S56 and L17 for the others.   
The template spectra have been selected on the basis of their 
spectral type (K and M), high S/N (40$-$200), 
low rotational velocities (line full width at half
maximum $<$0.6~\AA) and because they do not show any accretion signatures.
However, the H$\alpha$, other emission lines and telluric lines
have been excluded from the cross-correlated spectral range.
In order to determine the RVs relative to the solar system, we measured 
the centroid shifts of a selected group of lines in the template spectra, 
using the IRAF task RVIDLINES.

For the stars with a single main peak in the cross-correlation 
function and with all the RVs in agreement within 2$\sigma$, we
computed a mean RV as the weighted average of the different measurements.
The measured RVs are given in Tables~\ref{tab:mag_star} and
\ref{tab:star_lambda}
for the $\sigma$~Ori and the $\lambda$~Ori  cluster, respectively.
The other stars, classified as candidate binaries, are discussed in 
Sect.~\ref{par:binaries}. 

\subsection{Rotational velocities}
We were able to measure the rotational velocities (v$\sin i$) 
of 20 stars and to estimate a v$\sin i$ upper 
limits of all the remaining members identified 
in the two clusters (see Sect. \ref{par:memb})
and not classified as binary systems (see Sect. \ref{par:binaries}). 

Rotational velocities were measured
by cross-correlating the spectra of each star 
with the spectra of the stars L04 (for stars with spectral-type earlier 
than M2) and S88 (for stars with later spectral-type stars), 
using the IRAF task FXCOR.
Both stars used as templates show very narrow lines 
(full width half maximum --FWHM--$<$0.540 \AA) and have S/N $>$20. 
Specifically, v$\sin i$ is related to the FWHM of
the Gaussian which better describes the peak of the cross-correlation 
function by a calibration function.
In order to derive the calibration function, 
we first cross-correlated each template spectra with itself artificially 
broadened at different velocities using the
rotational profile of \cite{Gray1992book}; 
then, we fitted the relation between the values of FWHM
and the rotational velocities with a third order polynomial.
   
Errors on v$\sin i$ depend on the uncertainties in the 
cross-correlation process between the template and the stars spectra; these
have been estimated by varying 
the Gaussian fit parameters within reasonable ranges.
Moreover, taking into account the features of the template 
spectra and the uncertainties in our procedure, we
concluded that the lowest measurable rotational velocity is 
$\sim$17~km/s, which corresponds to a line broadening equal 
to the instrumental broadening.

Inferred v$\sin i$ and upper limits are listed in Col.~12 in
Tables~\ref{tab:mag_star} and
\ref{tab:star_lambda}.

\subsection{Pseudo equivalent width of absorption and emission lines
\label{par:pew}}

We measured the equivalent width of the Li line at 6708~\AA\ and of 
5 emission lines (H$\alpha$, NII at 6583~\AA, HeI at 6678~\AA\ and SII at 
6716 and 6731~\AA), using the IRAF task SPLOT, by integrating the area under 
the continuum level. For the lithium line, we integrated all the spectra
over the same range 
(6705.8$-$6709.5~\AA\ at a rest reference frame).
For the emission lines, we integrated on a spectral range different from
star to star, because 
accretion lines can be shifted or broadened, depending on the
velocity of the outflowing or accreting material.
The lithium line measurements are reported in Tables~\ref{tab:mag_star} and
\ref{tab:star_lambda}, while the 
emission line measurements are reported in Tables~\ref{tab:acc_sori} 
and \ref{tab:acc_lor}, for the $\sigma$~Ori and the $\lambda$~Ori clusters, 
respectively.
Due to the presence of molecular bands, which strongly affect the spectrum of 
late-type stars, the EWs are measured with respect to a false
continuum. To remark this point, in the rest of the paper we refer to these 
measurements as pseudo-equivalent widths (pEWs). 
We did not measure any pEW of the double-lined binaries and of the star L48, 
because in the former case we could not separate the contributions of the 
different components of the binary system, while in the latter the S/N 
is too low to estimate the continuum level after the sky subtraction.  

For the Li and H$\alpha$ lines, we derived the pEWs as the 
average of 3 independent measurements carried out by selecting a minimum, a
maximum and a median continuum level. Errors have been computed as the 
half-difference between the maximum and minimum of the 3 measurements. 
We measured the Li pEW by integrating over the 6705.8$-$6709.5~\AA\ range
even in the case where the presence of the Li line was not evident. In these
cases, the line 
pEW is always below 150~m\AA, while in the other cases the pEW
is always higher than 250~m\AA\ except for the star S55, which,  
as discussed in \cite{Sacco2007A&A}, has started to deplete its photospheric 
lithium.

For the pEWs of other emission lines, we took a single measurement
and, whenever the presence of the line was not evident in the spectrum, 
we estimated an upper limit on the basis of the faintest measurable lines
observable in the same spectral range.

\subsection{H$\alpha$ width at 10\% of the peak \label{par:ha_width}}

We measured the width of the H$\alpha$ line at 10\% of the peak in order to
discriminate between accreting and non-accreting objects and to derive the
mass accretion rates (MARs), using the following relationship:

\begin{equation}
\log (\dot M_{acc})=-12.89(\pm 0.3)+9.7(\pm 0.7)\times 10^{-3} 
H\alpha_{10\%}
\label{eqn:Natta}
\end{equation}

\noindent where H$\alpha_{10\%}$ is the H$\alpha$ width at 10\% of the peak
in km s$^{-1}$ and $\dot M_{acc}$ is the MAR in $M_{\odot}$~yr$^{-1}$.
This relationship was empirically derived by \cite{Natta2004A&A}, using 
independent measurements of MARs and  H$\alpha$ widths in a sample of 
stars spanning a range of masses from about 0.04 to 0.8~$M_{\odot}$.
According to \cite{Natta2004A&A}, this relationship 
holds for H$\alpha_{10\%}\geq 200$~km~s$^{-1}$, but other authors
\citep{White2003ApJ} used the more conservative threshold of 270~km~s$^{-1}$.

The H$\alpha$ widths of cluster members and MARs of the stars harboring a
disk are reported in Tables~\ref{tab:acc_sori} and \ref{tab:acc_lor} for
$\sigma$~Ori and
$\lambda$~Ori, respectively. Errors on the H$\alpha$ widths 
depend on the uncertainties in the continuum flux, which affect the 
determination of the 10\% level, while the errors on the MARs depend both on 
the errors on the H$\alpha$ width and on the uncertainties on the 
parameters included in Eq.~(1). For binary systems the H$\alpha$ widths have
been considered undetermined in order to avoid systematic errors related to
the presence of two spectra shifted at different velocities.  

\subsection{H$\alpha$ variability}

We investigated the variability of the H$\alpha$ pEWs for all the
stars, and of the H$\alpha$ width at 10\% of the peak for the CTTSs, by
comparing the measurements performed in each individual observing run. 
Among the WTTSs the median
variability ((max(pEW)-min(pEW))/average(pEW)) ranges between 0.1 and 1.9,
with a median value of 0.37, but it is larger than 1.0 only for three stars
(S28, L06, L36). For the latter stars, the variation is observed only in a
single run, suggesting that it was probably recorded during a flaring
event.

\begin{figure}
\resizebox{\hsize}{!}{
\includegraphics{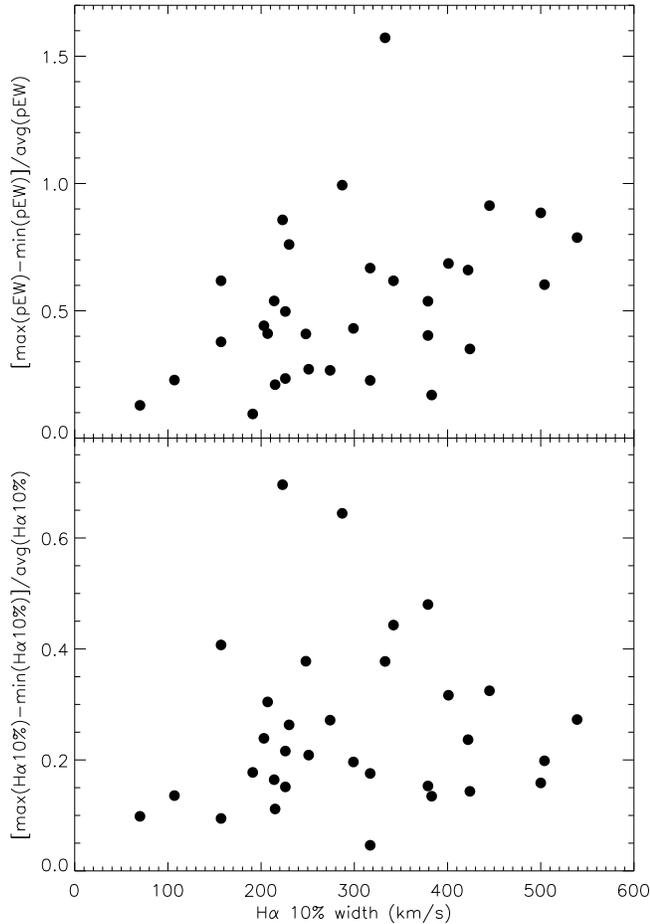}}
\caption{H$\alpha$ variability of the CTTSs as a function of the H$\alpha$
width. The upper and bottom panels
show, respectively, the variability of the pEWs and of the H$\alpha$ width at 
10\% of the peak.}
\label{fig:CTTS_var}
\end{figure}  

The variability of the pEWs and of the 10\% widths for the CTTSs
is plotted in Fig.~\ref{fig:CTTS_var} as a function of the 10\% widths measured
on the summed spectra. As shown in the figure, we do not find evidence for
any correlation between the two measurements of variability and the 10\% width,
which, as explained in the previous section, is correlated with the mass
accretion rate.  The median value, the mean and the standard deviation of
the variability of the pEWs are 0.50, 0.53 and 0.31, respectively, while the
median, mean and standard deviation of the 10\% widths are 0.22, 0.26 and
0.15, respectively. In 5 cases (S05, S33, S65, S98 and L34)
for the pEWs and in 4 cases (S12, S65, S69 and L34) for the 10\% widths,
the variations differ by more than 3$\sigma$ from the mean of the sample.
In the case of the star S23, which is not included in Fig.~\ref{fig:CTTS_var},
we find a variability of the H$\alpha$ pEW $\ge$4, because the line
is composed of both absorption and emission features, and the pEW is
positive in some runs
and negative in others.

Moreover, the intrinsic variability is in most cases higher than
the errors on the measurements of the H$\alpha$ pEWs and 10\% widths, which
are of the order of 10\%, as  can be seen from Tables~\ref{tab:acc_sori} and
\ref{tab:acc_lor}.

\section{Results}
\label{par:results}

\subsection{Binaries\label{par:binaries}}

\begin{figure*}
\resizebox{\hsize}{!}{\includegraphics{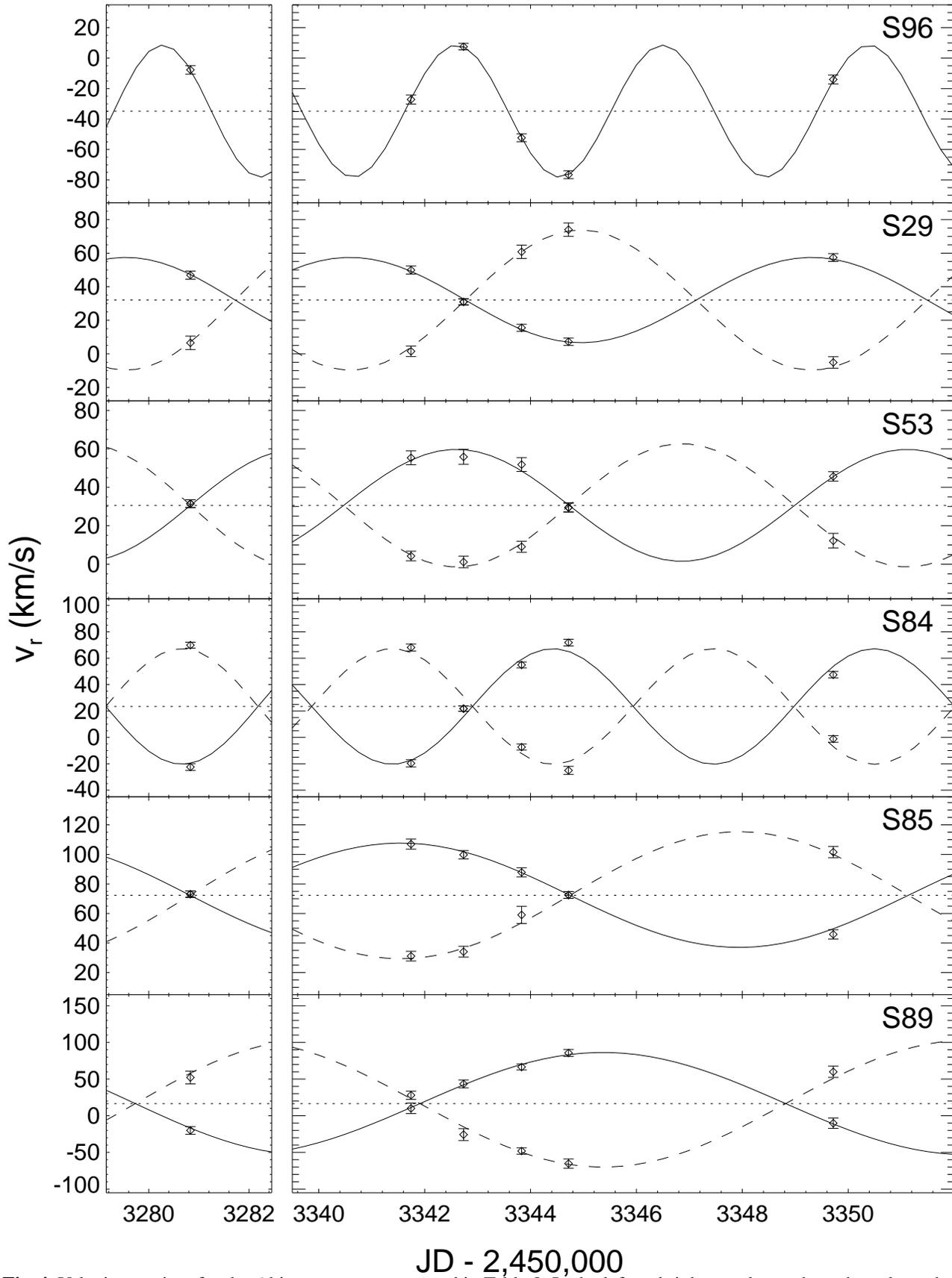}}
\caption{Velocity vs. time for the 6 binary systems reported
in Table~\ref{tab:bin}. In the left and right panels we show
the values obtained from the runs performed in October and December,
respectively. The continuous and dotted curves show
the sine functions which fit the data.}
\label {fig:fit_sb}
\end {figure*}

We found 11 stars in $\sigma$~Ori and 5 in 
$\lambda$~Ori for which
at least 2 of the measured RVs differ by more than $2\sigma$,
and therefore we classified them as candidate binaries.
The measured RVs in each run are given in Tables~\ref{tab:vr_bin_sori} and
\ref{tab:vr_bin_lori}. Specifically, in the $\sigma$~Ori sample, for 6
binaries
all the RVs are not in agreement among each other. Of these, 5 stars show a
cross-correlation function with two distinct peaks, corresponding to the
RVs of both components of the binary system. 
The RVs as a function of time for these 6 binaries 
are plotted in Fig.~\ref{fig:fit_sb}, and show a
well-defined sinusoidal trend.
We have performed a least-squares fit to the RV curves using a sine
function; in the case of double-lined binaries, both curves were fitted 
simultaneously. 

The binary system parameters derived from the fits with their
1$\sigma$ uncertainties are given in
Table~\ref{tab:bin}. For the other 5 
candidate binaries in $\sigma$~Ori, only
the RV measured in October 2004 does not agree with the other 5 RVs 
measured in
December 2004, that are compatible within the errors. A careful
check of the October spectra did not evidence any reduction or analysis
problem, therefore we believe that the observed discrepancy is real.
These stars might therefore be binaries with a
period of some months, for which RV variations cannot be detected 
in the observations taken in December, which are distributed over a
period shorter than 10 days. Further RV measurements are required to confirm 
their nature.

In the $\lambda$ Ori cluster, we found one star with two distinct peaks
in the
cross-correlation function, for which however we cannot obtain a fit to the
RV data, and 4 stars with at least two of the measured 
RVs differing by more than twice the error bar.

\setcounter{table}{5}
\begin {table*}[!ht]
\caption{\label{tab:vr_bin_sori} Radial velocities of the candidate binary systems identified in the
$\sigma$ Ori cluster sample.}
\centering
\begin{tabular}{ccccccc}
\hline 
\hline
 & \multicolumn{6}{c}{JD-2453000.00}\\
 & 280.82 & 341.73 & 342.73 & 343.82 & 344.71 & 349.71\\                   
\hline
ID & RV (km/s) & RV (km/s)  & RV (km/s)  & RV (km/s)  & RV (km/s)  & RV (km/s) \\                   

 \hline
  S04 & 17.83 $\pm$ 4.81 & 33.10 $\pm$ 3.01 & 34.19 $\pm$ 6.17 & 32.21 $\pm$
  4.64 & 27.55 $\pm$ 6.40 & 34.94 $\pm$ 4.57 \\  

 S26 & 27.95 $\pm$ 3.38 & 43.72 $\pm$ 2.90 & 43.07 $\pm$ 3.35 & 44.88 $\pm$
  3.34 & 45.25 $\pm$ 3.18 & 44.32 $\pm$ 4.01 \\  

 S29a & 6.56 $\pm$ 4.03 & 1.51 $\pm$ 3.13 & 
 30.97 $\pm$ 1.82 & 60.81 $\pm$ 3.99 &
 73.99 $\pm$ 3.96 & -5.13 $\pm$ 3.42 \\

S29b & 46.91 $\pm$ 2.40 & 49.93 $\pm$ 2.41 & 
 30.97 $\pm$ 1.82 & 15.51$\pm$ 2.11&
 7.20 $\pm$ 2.18 & 57.42 $\pm$ 2.32  \\

 S43 & 39.17 $\pm$ 1.41 & 21.10 $\pm$ 1.59 & 18.60 $\pm$ 1.67 & 20.96 $\pm$
  1.53 & 20.13 $\pm$ 1.62 & 22.40 $\pm$ 1.30 \\  

 S50 & 11.81 $\pm$ 1.13 & 4.43 $\pm$ 1.09 & 2.95 $\pm$ 1.34 & 5.41 $\pm$
  1.12 & 4.77 $\pm$ 1.69 & 6.18 $\pm$ 1.21 \\  

S53a & 31.50 $\pm$ 2.03 & 55.34 $\pm$ 3.59& 
 55.84 $\pm$ 3.90& 51.82 $\pm$ 3.58 &
 29.51 $\pm$ 2.34 & 45.68 $\pm$ 2.44  \\  

S53b & 31.50 $\pm$ 2.03 & 4.24 $\pm$ 2.52 & 
 1.14 $\pm$ 2.97 & 9.07$\pm$ 2.82 &
 29.51 $\pm$ 2.34 & 12.21 $\pm$ 3.82  \\


S84a & 69.90 $\pm$ 2.24 & 68.11 $\pm$ 2.64 & 
21.78 $\pm$ 2.06& -7.34 $\pm$ 2.29 &
 -25.03 $\pm$ 3.06& -1.15 $\pm$ 2.50  \\  

S84b & -22.45 $\pm$ 2.53 & -19.67 $\pm$ 2.75 & 
21.78 $\pm$ 2.06& 54.91$\pm$ 2.20 &
 71.85 $\pm$ 2.51 & 47.48 $\pm$ 2.47  \\

S85a & 73.12 $\pm$ 2.15& 31.13 $\pm$ 3.29 & 
34.13 $\pm$ 3.68 & 59.05 $\pm$ 5.85&
 72.54 $\pm$ 2.28 & 101.53 $\pm$ 3.80  \\  
 
S85b & 73.12 $\pm$ 2.15 & 106.96 $\pm$ 3.39 & 
99.75 $\pm$ 2.87 & 87.84$\pm$ 3.10 &
 72.54 $\pm$ 2.28 & 45.87 $\pm$ 3.13  \\

S89a & 52.14 $\pm$ 8.76 & 27.99 $\pm$ 5.64 & 
-25.75 $\pm$ 8.10& -48.08 $\pm$ 4.31 &
 -65.28 $\pm$ 6.26 & 59.86 $\pm$ 7.67  \\

S89b & -20.04 $\pm$ 5.23 & 10.07 $\pm$ 7.17 & 
43.42 $\pm$ 5.42& 66.46$\pm$ 4.08 &
 85.63 $\pm$ 4.75 &-10.17 $\pm$ 7.02  \\

 S91 & 75.36 $\pm$ 1.44 & 81.10 $\pm$ 2.27 & 81.19 $\pm$ 1.91 & 83.45 $\pm$
  1.61 & 83.22 $\pm$ 1.91 & 84.46 $\pm$ 1.62 \\  

S96 & -7.70 $\pm$ 2.77 & -27.13 $\pm$ 2.98 & 7.54 $\pm$ 2.12 & -52.30 $\pm$
  2.54 & -76.51 $\pm$ 2.62 & -13.99 $\pm$ 2.95 \\

\hline
 \end{tabular}
\end{table*}

\setcounter{table}{6}
\begin {table*}[!ht]
\caption{\label{tab:vr_bin_lori} Radial velocities of the candidate binary systems identified in the
$\lambda$ Ori cluster sample.}
\centering
\begin{tabular}{ccccccccc}
\hline
\hline
 & \multicolumn{8}{c}{JD-2453600.00}\\ 
   & 58.86 &  59.86 & 60.84 & 84.82 & 86.78  & 86.82 & 86.85 & 97.74\\ 
\hline
ID & RV (km/s) & RV (km/s)  & RV (km/s)  & RV(km/s)  & RV (km/s)  & RV (km/s) & RV (km/s) & RV (km/s) \\ 
 \hline
L01  &  25.41$\pm$1.06 &  29.09$\pm$1.25 &  24.36$\pm$1.05 &  25.65$\pm$0.97 &  24.05$\pm$0.74 &  25.03$\pm$1.16 &  24.98$\pm$0.91 &  24.31$\pm$1.46 \\
L03  &  19.48$\pm$0.41 &  20.98$\pm$0.49 &  20.13$\pm$0.37 &  20.91$\pm$0.41 &  20.89$\pm$0.54 &  20.91$\pm$0.75 &  21.01$\pm$0.62 &  32.16$\pm$0.92 \\
L10a &  27.18$\pm$0.66 &  29.27$\pm$1.44 &  28.14$\pm$0.73 &  -1.29$\pm$1.88 &  -2.48$\pm$1.90 &  -1.66$\pm$2.06 &  -1.33$\pm$1.90 &  26.59$\pm$0.66 \\
L10b &  27.18$\pm$0.66 &  29.27$\pm$1.44 &  28.14$\pm$0.73 &  30.49$\pm$2.80 &  29.34$\pm$2.99 &  29.71$\pm$2.57 &  29.19$\pm$2.58 &  26.59$\pm$0.66 \\
L22  &  24.40$\pm$1.43 &  28.26$\pm$2.58 &  29.54$\pm$1.31 &  24.54$\pm$1.67 &  27.19$\pm$1.15 &  27.57$\pm$0.97 &  27.60$\pm$0.88 &  23.27$\pm$1.33 \\
L32  &  25.41$\pm$2.88 &  24.66$\pm$3.28 &  16.12$\pm$3.85 &  28.19$\pm$2.63 &  25.35$\pm$2.98 &  24.67$\pm$3.96 &  25.16$\pm$2.67 &  34.81$\pm$5.05 \\
\hline
\end{tabular}
\end{table*}

\setcounter{table}{7}
\begin{table*}[!ht]
\caption{\label{tab:bin} Binary systems parameters for the 6 
binaries in the $\sigma$~Ori cluster for which a sinusoidal fit 
of the RV curves was performed. Errors are $1\sigma$}
\begin{center}
\begin{tabular}{lrrccrrcc}
\hline
\hline
ID& $V_\circ^{\,a}$ (km/s)& P$^{\,b}$ (days)& $K_1^{\,c}$ (km/s)& 
$K_2^{\,c}$ (km/s)& $a_1 \sin i^{\,d}$ ($R_\odot$)& 
$a_2 \sin i^{\,d}$ ($R_\odot$)& f(M)$^{\,e}$ ($M_\odot$)& 
$M_1/M_2^{\,f}$ \\
\hline
S29&  32.06 $\pm$ 0.73  &  8.72 $\pm$ 0.02 & 25.4 $\pm$ 1.3& 41.7 $\pm$ 1.3& 
 4.38 $\pm$ 0.22&  7.18 $\pm$ 0.22& 0.015 $\pm$ 0.002& 1.64 $\pm$ 0.10\\
S53&  30.61 $\pm$ 1.06  &  8.52 $\pm$ 0.01 & 29.1 $\pm$ 2.8& 32.1 $\pm$ 2.3& 
 4.90 $\pm$ 0.46&  5.39 $\pm$ 0.39& 0.022 $\pm$ 0.006& 1.10 $\pm$ 0.13\\
S84&  23.43 $\pm$ 0.69  &  6.07 $\pm$ 0.01 & 43.8 $\pm$ 1.3& 43.8 $\pm$ 1.3& 
 5.25 $\pm$ 0.16&  5.25 $\pm$ 0.16& 0.053 $\pm$ 0.005& 1.00 $\pm$ 0.04\\
S85&  72.34 $\pm$ 0.88  & 12.78 $\pm$ 0.02 & 35.3 $\pm$ 2.3& 43.0 $\pm$ 2.3& 
 8.91 $\pm$ 0.59& 10.84 $\pm$ 0.59& 0.058 $\pm$ 0.012& 1.22 $\pm$ 0.10\\
S89&  16.54 $\pm$ 2.02  & 13.82 $\pm$ 0.04 & 69.8 $\pm$ 4.4& 86.5 $\pm$ 4.4& 
19.02 $\pm$ 1.19& 23.59 $\pm$ 1.20& 0.486 $\pm$ 0.092& 1.24 $\pm$ 0.10\\
S96& $-$34.79 $\pm$ 1.19& 3.895 $\pm$ 0.004& 43.4 $\pm$ 1.6& \ldots& 
 3.33 $\pm$ 0.12& \multicolumn{1}{c}{\ldots}& 0.033 $\pm$ 0.004& \ldots \\
\\
\hline
\multicolumn{9}{l}{$a$: velocity of the binary system center of mass} \\
\multicolumn{9}{l}{$b$: period of the binary system} \\
\multicolumn{9}{l}{$c$: semi-amplitude of the RV curve}\\
\multicolumn{9}{l}{$d$: semi-major axis of the binary orbit}\\
\multicolumn{9}{l}{$e$: binary system mass function}\\
\multicolumn{9}{l}{$f$: mass ratio of the binary components}\\
\end{tabular}
\end{center}
\end{table*}

For the selection of members from RVs, we used the derived center of
mass velocities for the binary systems listed in Table~\ref{tab:bin}. For
the other binaries, the system velocities cannot be determined, and will
therefore not be used for their membership determination.

\subsection{Membership\label{par:memb}}

\begin{figure}
\resizebox{\hsize}{!}{
\includegraphics{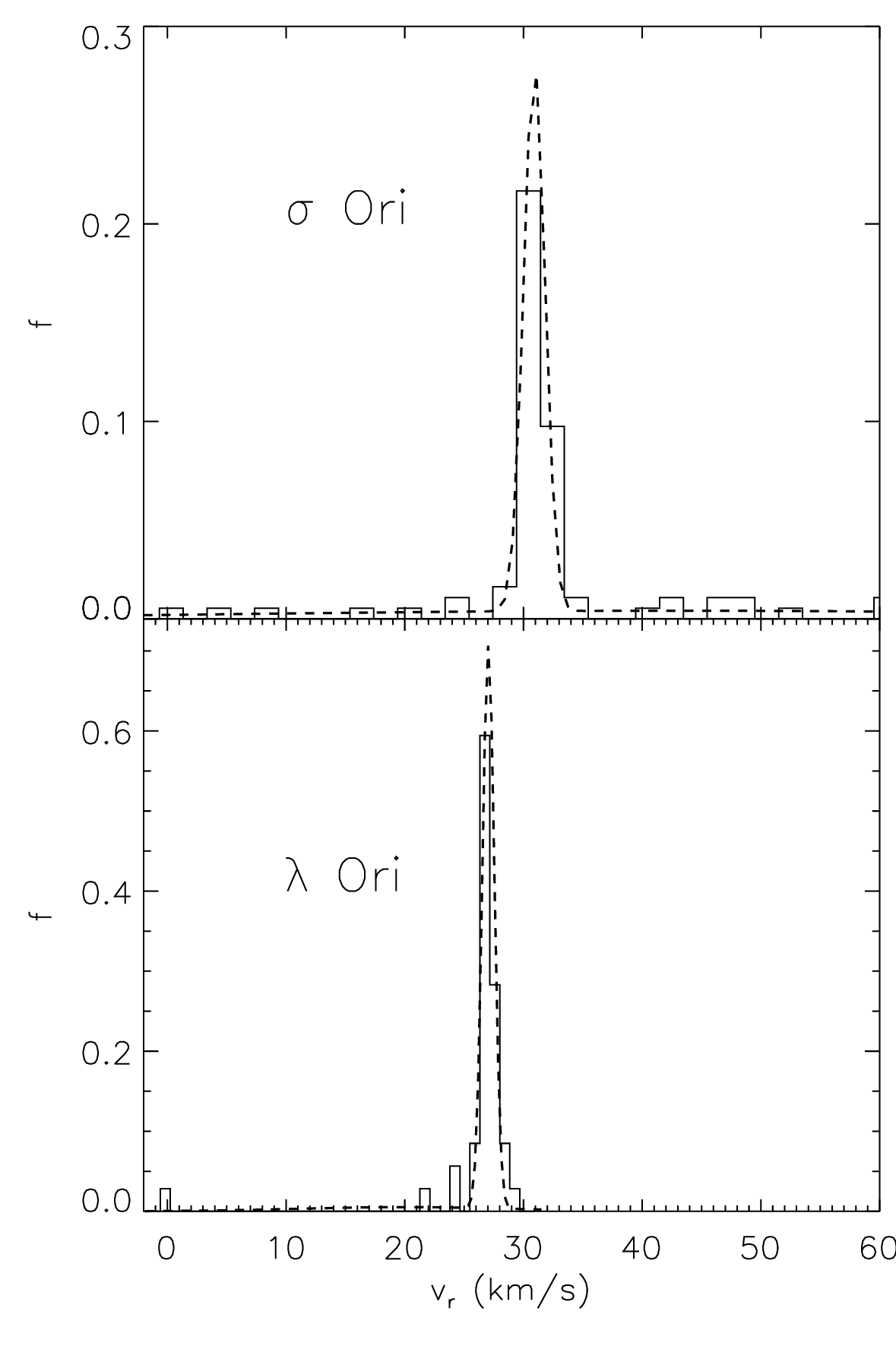}}
\caption{Radial velocity distribution of the observed sources in the two
clusters. Only single stars and the binaries of Table~\ref{tab:bin} are
included. The dashed curves are the functions resulting from the fit of the 
velocity distribution with a weighted sum of two Gaussians.}
\label{fig:gaussian}
\end{figure}  

To determine cluster membership we used three independent criteria,
based on the RV distribution, the pEW of the Li absorption line, and the
presence of the H$\alpha$ line in emission.

In Fig.~\ref{fig:gaussian} we show the RV distributions for the observed
stars in $\sigma$~Ori and $\lambda$~Ori, excluding the 10 probable binaries
(5 for each cluster) for which the velocity of the center of mass is
undetermined. The two distributions have been fitted with the weighted sum
of two Gaussian functions (shown in Fig.~\ref{fig:gaussian} with dashed
curves), one describing the velocity distribution of the cluster members and
the other describing the velocity distribution of the field stars. For
$\sigma$~Ori the cluster distribution is centered at $V_\mathrm{C} =
30.93$~km~s$^{-1}$, with a standard deviation $\sigma_\mathrm{C} =
0.92$~km~s$^{-1}$, while for the field star distribution we find
$V_\mathrm{F} = 43$~km~s$^{-1}$ and $\sigma_\mathrm{F} = 37$~km~s$^{-1}$. In
the case of $\lambda$~Ori we obtain $V_\mathrm{C}=27.03$~km~s$^{-1}$ with
$\sigma_\mathrm{C}=0.49$~km~s$^{-1}$ for the cluster and $V_\mathrm{F} =
22$~km~s$^{-1}$ with $\sigma_\mathrm{F} =9$~km~s$^{-1}$ for the field,
although, considering the low number of field stars in the $\lambda$~Ori
sample, the parameters of the latter distribution are poorly determined. For
each cluster, we classify a star as member if its RV, taking the error bar
into account, differs by less than 3$\sigma_\mathrm{C}$ from the cluster
average velocity $V_\mathrm{C}$. Using this criterion, among the 93 stars in
the $\sigma$~Ori sample we find 64 stars with RV consistent with membership,
while among the 44 stars in the $\lambda$~Ori sample, 37 have RV consistent
with membership. The membership of the 10 binaries excluded from the RV
distributions is undetermined. The expected contamination of the $\sigma$
Ori member sample by non-members, estimated by integrating the field star
distribution between $V_\mathrm{C}-3\sigma_\mathrm{C}$ and $V_\mathrm{C} +
3\sigma_\mathrm{C}$, is $\sim$2 stars.

The velocity of the $\sigma$~Ori cluster is in agreement with that
of its central high-mass star $\sigma$~Orionis ($27\pm 4$~km~s$^{-1}$,
\citealt{Morrell1991ApJS}) and with the cluster mean velocity found by
\cite{Zapatero2002A&A}, \cite{Kenyon2005MNRAS} and \cite{Jeffries2006MNRAS}.
Our measurements do not evidence the presence of the kinematically-separated
($V_r=23.8\pm 1.1$~km~s$^{-1}$) young stellar population discovered by
\cite{Jeffries2006MNRAS}. Since this second population is concentrated to the
north-west of the hot star, this apparent disagreement can be attributed to
the different positions of the observed fields. In fact, if we limit the
analysis to the \cite{Jeffries2006MNRAS} field with the largest area in
common with our field, and compute the number of stars in the two RV ranges
(20-27 and 27-35 km/s) defined by \cite{Jeffries2006MNRAS} for the two
populations, we find in our field a ratio (3/65 or 5$\pm$3\%) consistent
with that found by \cite{Jeffries2006MNRAS} (4/46 or 9$\pm$4\%). The
velocity of the $\lambda$~Ori cluster is in agreement with that found by
\cite{Dolan1999AJ} ($24.3\pm 2.8$~km~s$^{-1}$), but slightly lower than
the velocity of its high-mass central star $\lambda$~Orionis ($30.10\pm
1.30$~km~s$^{-1}$, \citealt{Kharchenko2007AN}).

\begin{figure}
\resizebox{\hsize}{!}{
\includegraphics{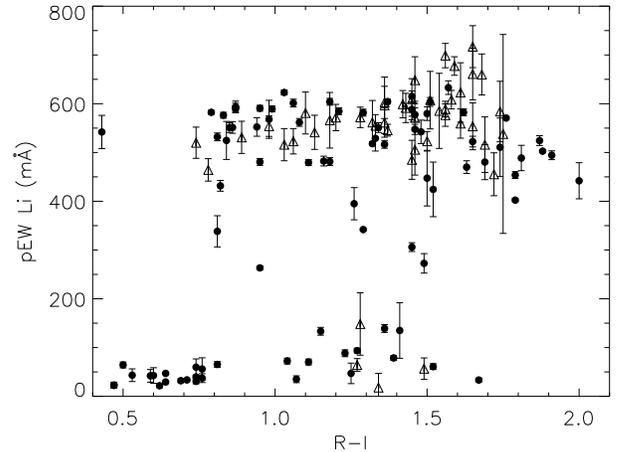} }
\caption{Li pEWs vs. $R-I$ for the stars 
of the $\sigma$~Ori (filled circles) and $\lambda$~Ori (open triangles) 
cluster samples with available optical photometry.}
\label {fig:pew_col}
\end {figure}

Late-type stars ($0.08-0.5 M_{\odot}$) deplete their photospheric lithium 
during the pre-main sequence (PMS) phase \citep{Bodenheimer1965ApJ, Siess2000A&A}, therefore the
presence of strong Li absorption at 6708~\AA\ in young clusters can be
used as an independent
criterion to identify cluster members. Figure~\ref{fig:pew_col} shows the Li
pEWs measured in the two clusters as a function of the $R-I$ color.
Since in all cases where the Li line was not clearly identified the 
Li pEW turned out to be less than 250~m\AA, we fixed the threshold for
selecting cluster members at this value. For the double-lined binaries and
for the star L48, for which we cannot measure the Li pEW, we 
established the membership only by considering whether the Li absorption
feature can be identified or not.
As shown in Fig.~\ref{fig:pew_col}, the bulk of the measured pEWs above the
chosen threshold is located around a median value of 560~m\AA\ with a
dispersion of $\sim$100~m\AA. According to the curves of growth
derived by 
\cite{Palla2007ApJ} for a sample of stars observed by GIRAFFE and very
similar to that described in this paper, this median pEW is compatible with
the full preservation of lithium. In the $\sigma$~Ori cluster, we found a
few sources with pEWs between 250 and 400~m\AA. As shown by
\citet{Sacco2007A&A},
these low pEWs are mainly due to spectral veiling, but we cannot exclude the
presence of some partially-depleted stars. This hypothesis is supported by
the discovery of three highly-depleted stars among the $\sigma$~Ori cluster
members, reported by \citet{Sacco2007A&A}.

Young K and M stars show emission in H$\alpha$ due to accretion and
chromospheric activity \citep{Barrado2003AJ}. Therefore we used 
the presence of H$\alpha$ in emission as the third criterion to identify
cluster members. Although, due to the very low S/N, we have not 
been able to subtract the correct emission of the sky from the spectrum of the star
L48 (see Sect.~\ref{par:pew}), 
this star has been classified as a member, considering 
that the H$\alpha$ emission line observed in its spectrum is nearly two times
larger than the H$\alpha$ line observed in the sky spectrum.

In Tables~\ref{tab:mag_star} and \ref{tab:star_lambda},
for the $\sigma$~Ori and $\lambda$~Ori clusters respectively, we give 
the results of our membership selection from the individual criteria
mentioned above, together with a final membership assessment derived 
from the comparison of the three criteria, which in most cases agree.
In Table~\ref{tab:mag_star} we also indicate if the X-ray counterparts of
the stars have been detected by \textit{XMM-Newton}.
For the probable binaries with undetermined RV, we relied only on the Li and
H$\alpha$ criteria, considering as possible members those stars for which
both Li and H$\alpha$ indicate membership, although we cannot exclude their
belonging to a separate young population present in the same area. 
In the $\sigma$~Ori sample, we classify 62 stars as members or possible
members and 29 stars as non-members. In $\lambda$~Ori, we find 42 members or
possible members and 2 non-members. For the remaining 7 stars in
$\sigma$~Ori and 5 stars in $\lambda$~Ori, the three criteria do not agree,
and we assigned a final membership based on the following considerations:

\begin{itemize}
\item star S37 in $\sigma$~Ori has a RV inconsistent with membership, but it
appears to be a PMS star both for H$\alpha$ and Li, as well as for the
presence of X-ray emission, which supports its youth. Its RV
($V_\mathrm{r}=23.73\pm 0.45$~km~s$^{-1}$) is similar to that of the second
population discovered by \citet{Jeffries2006MNRAS} close to the $\sigma$~Ori
cluster ($V_\mathrm{r}=23.8\pm 1.1$~km~s$^{-1}$), therefore this star might
belong to this separate population in the Orion OB1 association. On the
other hand, \citet{Zapatero2002A&A} found for S37 a RV of $33\pm
7$~km~s$^{-1}$, which is not consistent with the value measured by us,
suggesting that this star might be a binary system. 
Although this second hypothesis cannot be excluded, the excellent agreement
of our precise RV measurement with the RV of the second population from
\citet{Jeffries2006MNRAS} and the absence of any sign of binarity in our 
observations lead us to believe the first hypothesis more likely, therefore
we classify 
this star as a probable non-member.

\item in $\lambda$~Ori, stars L11, L12 and L31 have RV inconsistent
with
membership, but H$\alpha$ and Li consistent with them being PMS stars. In
the case of L12, \citet{Dolan1999AJ} found a different RV of
30.52~km~s$^{-1}$; considering that no other young stellar population has
been discovered in the region of the $\lambda$~Ori cluster, this star is
likely a binary, therefore it has been classified as a possible member. We
classified also L11 and L31 as a possible cluster members,
considering that their RV differs from the mean velocity of the cluster by
less than 5$\sigma$.

\item the three stars S84, L05 and L13 are not members according to RV
and Li, but show H$\alpha$ in emission. They probably are older
chromospherically active field stars, therefore they are classified as 
non-members. 

\item stars S22 and S31 have RVs consistent with membership, but show 
H$\alpha$ in absorption and no signs of Li absorption, so we have 
classified these sources as non-members; 

\item stars S55, S59 and S75 are cluster members according to
RV and H$\alpha$, but their Li pEWs are below the threshold chosen for the
membership selection. As discussed in detail by \citet{Sacco2007A&A}, these
3 stars can be classified as cluster members also on the basis of other
membership indicators. 

\end{itemize}

\begin{figure*}
\centering
\resizebox{\hsize}{!}{\includegraphics{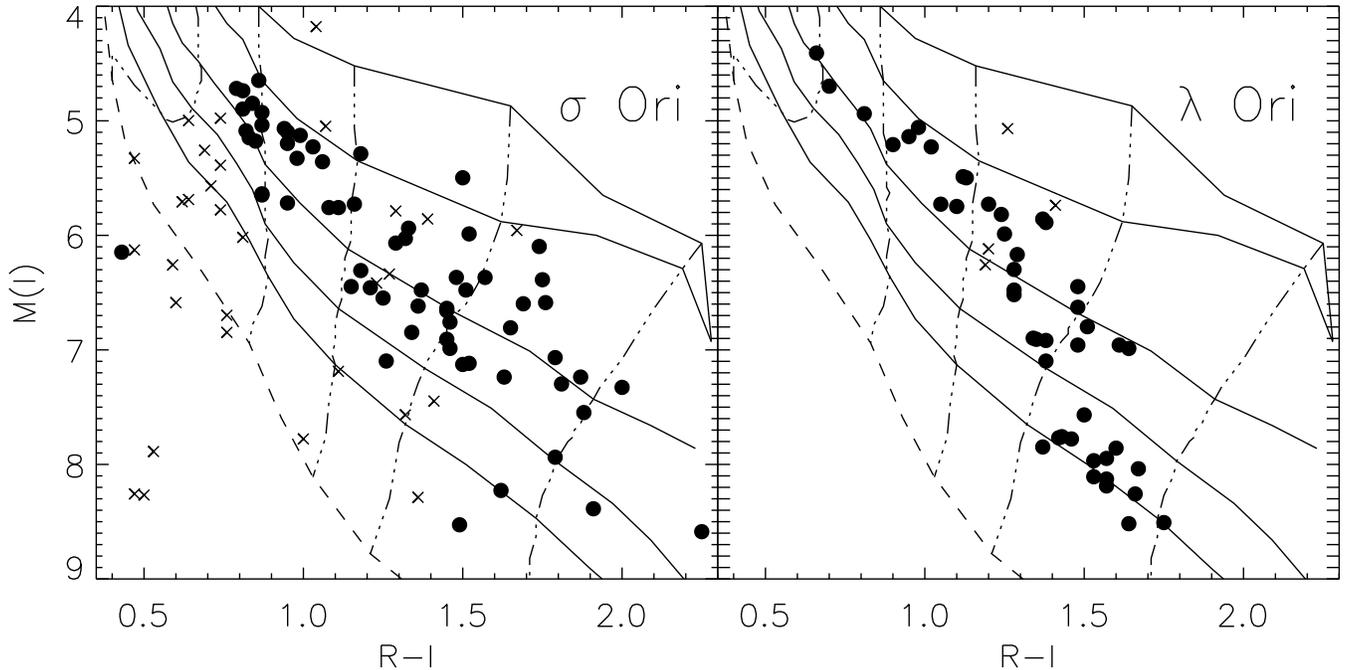} }
\caption{CMD of the observed targets in the $\sigma$~Ori (left panel)
and $\lambda$~Ori (right panel) cluster. Filled circles and crosses
indicate, respectively,  
members and non-members of the two clusters.
Stellar magnitudes are corrected for the distance
and reddening as in Fig.~\ref{fig:targetCMD}. Continuous,
dashed and dot-dashed lines represent the isochrones at 1, 2, 5, 10 and 20 Myr,
the Zero Age Main Sequence (ZAMS) and evolutionary tracks at 0.2, 0.3, 0.4,
0.6, 1.0 $M_{\sun}$ from the \cite{Siess2000A&A} models, respectively.
}
\label{fig:membCMD}
\end{figure*}

In conclusion, in the $\sigma$~Ori sample 
we find 65 members or possible members and 33 non-members, while in the
$\lambda$~Ori sample we find 45 members or possible members and 4
non-members. The higher contamination found in the $\sigma$~Ori sample can
be ascribed to the target selection method. Indeed, as shown in the color-magnitude 
diagram (CMD)
plotted in Fig.~\ref{fig:membCMD}, the $\sigma$~Ori sample includes a large
number of stars located below the 10$-$20~Myr isochrones, nearly all of
which turned out to be non-members. Stars below the 10$-$20~Myr isochrones 
are not included in the \cite{Barrado2004ApJ} and \cite{Dolan1999AJ}
catalogues, from which we retrieved the $\lambda$~Ori targets.

We have compared our results with the previous membership information
available in the literature (see Sect.~\ref{par:T_selection}). In the case of
$\sigma$~Ori, among the 18 members selected on the basis of previous
spectroscopic information, we confirm membership for all of them, except for
the star S37, which has been classified as a non-member because of its RV (see
above). Of the 66 candidates selected only by means of the photometric
data, only 47 are confirmed spectroscopically as members, while the
remaining 19 resulted to be non-members. For the remaining 14 targets which
appear to be non-members on the basis of photometry, we confirm that
13 of them are indeed not members of the cluster. The remaining star (S47) 
has been classified by us as a member, because it has all spectroscopic
indicators consistent with membership, and is also an X-ray source.
However its position in the CMD is anomalous, falling below the ZAMS
(see Fig.~\ref{fig:membCMD}). Given the spectroscopic and X-ray evidence,
we believe that its photometric data, retrieved from \cite{Wolk1996PhDT},
might be wrong, therefore its photometry
will not be taken into account in the following sections.
 
For $\lambda$~Ori, we find that all the targets selected by means of
previous spectroscopy from \cite{Dolan1999AJ} (15 stars) and
\cite{Barrado2004ApJ} (3 stars) are confirmed as members, while among the
candidates selected only by means of photometry we find 26 members or
probable members and 4 non-members.

\subsection{Accretion properties \label{par:acc}}

In Tables~\ref{tab:acc_sori} and \ref{tab:acc_lor}  we report, for
$\sigma$~Ori and $\lambda$~Ori respectively, all data concerning accretion,
outflow and disk signatures, namely, the H$\alpha$ width at 10\% of the
peak, the pEWs of H$\alpha$ and the pEWs of the HeI emission lines,
generally 
interpreted as due to the accretion flow, the pEWs of the forbidden NII 
and SII lines at 6583, 6716 and 6731~\AA, which are 
signatures of outflow material and, in the last column, the 
\textit{Spitzer} classification performed by \cite{Hernandez2007ApJ}  
and \cite{Barrado2007ApJ}, using the criterion defined by \cite{Lada2006AJ} 
based on the slope $\alpha$ of the spectral energy distribution 
between 3.6 and 8.0 $\mu$m, where 
$\alpha=\mathrm{d}\log(\lambda F_{\lambda})/\mathrm{d}\log(\lambda)$\footnote{Stars with
$\alpha >0$ are 
classified as protostars (class I), those with $-1.8<\alpha <0$  
as objects with a thick disk (class II), those with $-2.56<\alpha <-1.8$ 
as objects with a thin and evolved circumstellar disk (EV) and those with 
$\alpha <-2.56$ are classified as diskless objects (class III)}.    
In Tables~\ref{tab:acc_sori} and \ref{tab:acc_lor} we also give the accretion 
rates derived from Eq.~(1) for every star harboring a circumstellar disk 
with an H$\alpha$ width at 10\% of the peak larger than 200 km~s$^{-1}$. 
In the $\sigma$~Ori cluster, MARs range between $10^{-11}
M_{\odot}$~yr$^{-1}$ and 
$10^{-7.7} M_{\odot}$~yr$^{-1}$, while in the $\lambda$~Ori cluster they
range between 
$10^{-11} M_{\odot}$~yr$^{-1}$ and $10^{-10.1} M_{\odot}$~yr$^{-1}$.

\begin{figure*}
\centering
\resizebox{\hsize}{!}{\includegraphics{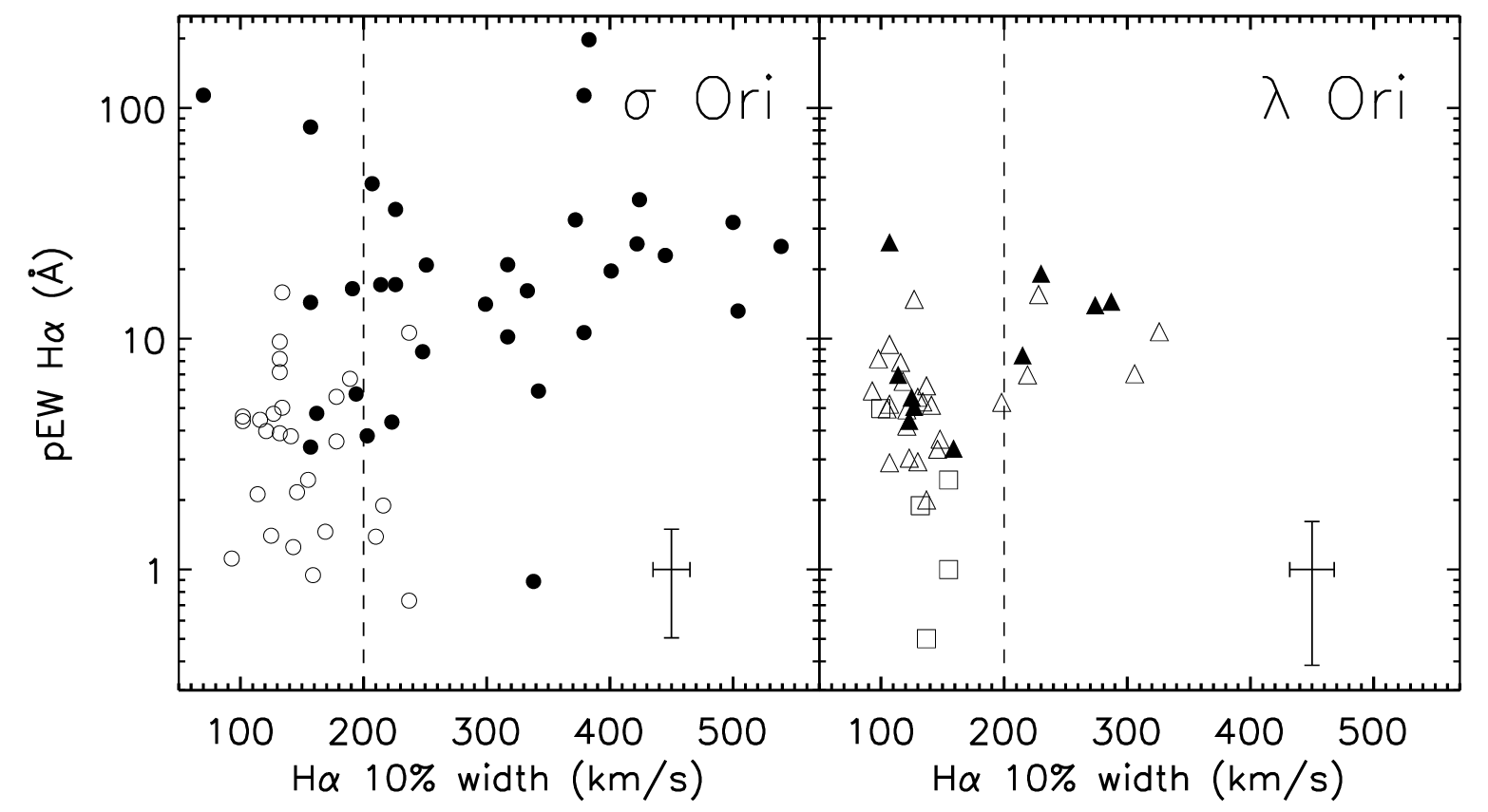}}
\caption{H$\alpha$ pEW as a function of the H$\alpha$ width at 10\% of the 
peak measured from the summed spectra of the $\sigma$~Ori (left panel) and 
$\lambda$~Ori (right panel) cluster members. Filled and open symbols represent stars with and without a
circumstellar disk according 
to the \textit{Spitzer} data, while squares in the left panel represent stars 
without \textit{Spitzer} photometry. Median error bars are reported on the
bottom-right corner.  
}
\label{fig:Ha_width}
\end{figure*}

The H$\alpha$ pEWs as a function of the H$\alpha$ widths at 10\% of 
the peak are shown in Fig.~\ref{fig:Ha_width}, where stars harboring
a circumstellar disk are indicated by filled symbols. 
As can be seen in this figure and in Tables~\ref{tab:acc_sori} and
\ref{tab:acc_lor}, we find some stars with an H$\alpha$ larger than
200~km~s$^{-1}$
without any other accretion signature and classified as class III objects by
the \textit{Spitzer} data. This result is likely due
to the 
presence of absorption bands in the H$\alpha$ spectral region, which causes
an overestimate of the H$\alpha$ width, and to the rotational broadening (the stars L16 and L33, which 
are the most relevant cases have v $\sin i \sim$ 60 km/s).
We also find some stars with H$\alpha$ larger than 200~km~s$^{-1}$ but 
with a low pEW (S23 and S47 is the most relevant case), and stars in the
opposite situation
(S93 and S94 are the most relevant cases). The former result is mainly due
to the
presence of intense absorption features overlapping 
the emission ones as already pointed out by \cite{Sicilia2006ApJ}
for some CTTSs belonging to the Tr37 cluster, while the latter result is due 
to the presence of accretion flows only in the direction perpendicular to the 
line of sight. 

We also observed the $\lambda$~Ori sources in the spectral range including 
the CaII infrared triplet, which can also be used to measure MARs 
\citep{Mohanty2005ApJ}, but in the whole sample we do not find any star clearly 
showing CaII emission due to accretion.

\section{Discussion}
\label{par:discussion}

\subsection{Disk properties in $\sigma$~Ori and $\lambda$~Ori \label{par:Spitzer}}

As mentioned in Sect.~\ref{par:T_selection} in the $\sigma$~Ori sample there
are 
78 sources in common with the \cite{Hernandez2007ApJ} catalogue of cluster
members and 5 sources in common with the catalogue of uncertain members.
According to our membership analysis, in the former catalogue there are 62
members and 16 non-members, while in the latter one there are 3 members 
and 2 non-members. Given that 20\% of the stars in common with our catalogue
are non-members, the number of $\sigma$~Ori members included in the
\cite{Hernandez2007ApJ} catalogue, selected using mainly photometric data,
is likely overestimated, and, therefore,
the fraction of T Tauri stars ($0.1-1.0 M_{\sun}$) with disks
(36$\pm$4\%) derived by them is 
likely underestimated. In the $\lambda$~Ori cluster, we have 44 sources in
common with the \cite{Barrado2007ApJ} catalogue, 40 classified as members
and 4 as non-members according to our analysis.

According to the disk classification by \cite{Hernandez2007ApJ} and
\cite{Barrado2007ApJ}, among the $\sigma$~Ori cluster members we have 1
protostar, 28 
stars with thick disks, 5 stars with evolved disks and 31 diskless stars,  
while in $\lambda$~Ori we have 7 stars with thick disks, 4 with evolved
disks and 
29 diskless stars. Therefore, in the mass range 0.2-1.0 $M_{\sun}$,
the fraction of stars with disks 
in the $\sigma$~Ori cluster (52$\pm$9\%) is larger than in the $\lambda$~Ori 
one (28$\pm$8\%).

The discrepancy between the two clusters does not depend on the analysis
method, because we compare the two clusters using homogeneous data 
and the same selection criteria both for membership and disk classification.
We can also exclude that this result might be due to a bias in the target
selection against the diskless stars. In fact, for both clusters the
observed stars have been retrieved from catalogues of candidate members
selected through CMDs   
and spectroscopic membership indicators (lithium and RVs), which do not 
depend on disk and accretion properties. 

Another possible bias in the target selection can be present against the
older stars, which are more likely class III objects. However, in contrast
with our results, this kind of bias should affect more strongly the
$\lambda$~Ori sample than the $\sigma$~Ori one. In fact, as pointed out in
Sect.~\ref{par:T_selection} and
\ref{par:memb}, the $\lambda$~Ori targets are selected mainly from
the \cite{Barrado2004ApJ} catalogue which includes only
stars around the 5 Myr isochrone (\citealt{Baraffe1998A&A} models at 400~pc), 
while the $\sigma$~Ori target sample includes stars in a larger age range
(0$-$20~Myr).
Moreover, the agreement between the fraction of stars with disks 
in our sample (28$\pm$8\%) and in the sample of all stars in 
the magnitude range 11.3$<J<$14.8 (105 stars) included 
in the \cite{Barrado2007ApJ} catalogue (22$\pm$5\%) 
confirms that our sample is not affected by target selection biases.
On the other hand, the disagreement between the fraction of stars with disks
found in 
the $\sigma$~Ori sample (52$\pm$9\%) and that obtained, in a 
very similar magnitude range (11.5$<J<$14.6) by \cite{Hernandez2007ApJ}
(36.3$\pm$4.1\%), 
confirms that, as we argued above, their sample of  cluster members
is contaminated by a large number of field stars.

The fraction of stars with disks found by us in $\sigma$~Ori is in
agreement with that found by \cite{Caballero2007A&Ab} for the brown dwarfs
(47$\pm$15\%). This suggests that the disk frequency does not depend on mass
over the interval 0.02$-$1.0~$M_{\sun}$.

\subsection{Accretion properties in $\sigma$~Ori and $\lambda$~Ori\label{par:acc_sori_lori}}

\setcounter{table}{8}
\begin {table*}[htb]
\caption{\label{tab:sori_vs_lori} Disk and accretion properties$^{\,a}$ in
$\sigma$ Ori and $\lambda$ Ori}
\centering
\begin{tabular}{cccccccc}
\hline 
 & Disk & Thick & H$\alpha$ & HeI  & other & active & active  \\   
 & & Disk &   width$^{\,b}$ &  6678 \AA & lines & disk (H$\alpha$)$^{\,c}$ &disk (He I)\\   
\hline
 \hline
$\sigma$ Ori  & 52$\pm$ 9\%  & 45$\pm$ 8\%  & 42$\pm$8\%  & 35$\pm$7\% & 37$\pm$8\% & 78$\pm$16\% & 65$\pm$14\%\\
              & (34/65)      &    (29/65)   & (25/60)      &  (22/63)    & (23/63) &   (25/32)    & (22/34) \\
$\lambda$ Ori & 28$\pm$ 8\%  & 18$\pm$ 7\%  &  10$\pm$5\% &  7$\pm$4\% & 12$\pm$5\% &  40$\pm$20\% & 27$\pm$16\% \\
              & (11/40)      &    (7/40)    &   (4/39)      &  (3/43)     & (5/43) &  (4/10)       & (3/11)   \\
 \hline
\multicolumn{7}{l}{$a$: Errors are calculated as $\sqrt{N}$/total.}\\ 
\multicolumn{7}{l}{$b$:  H$\alpha_{10\%}>$200 km/s}\\
\multicolumn{7}{l}{$c$: Stars harbouring a circumstellar disk with H$\alpha_{10\%}>$200 km/s}\\
\multicolumn{7}{l}{$c$: Stars harbouring a circumstellar disk with He I in emission}\\
\end{tabular}
\end{table*}

In Table~\ref{tab:sori_vs_lori} we compare the disk and accretion properties
of $\sigma$~Ori and $\lambda$~Ori on the basis of all the different disk and
accretion classifications. Furthermore, in the last two columns, we
compare the fraction of active disks, namely, the fraction of accreting
objects among the stars harboring a circumstellar disk (class I, class II
and EV).

The data reported in Table~\ref{tab:sori_vs_lori}, together with the
comparison between the accretion rates measured by means of the H$\alpha$
widths at 10\% of the peak, prove that, in spite of the similarities between
the two clusters,
the disk and accretion properties of their low-mass stellar populations are
very different. Specifically, in the $\sigma$~Ori cluster we find a larger
fraction of stars with disks, a larger fraction of accreting objects, and
larger accretion rates than in $\lambda$~Ori. Furthermore, the last two
columns of Table~\ref{tab:sori_vs_lori} show that the two clusters
differ also if we consider only the subsample of stars with a circumstellar
disk.

\subsection{The effects of the high-mass stars 
and of the supernova explosion in the $\lambda$~Ori cluster.}

\cite{Dolan1999AJ, Dolan2001AJ} already evidenced the lack of
strong H$\alpha$ emitters in the $\lambda$~Ori cluster and the existence 
of a discrepancy with the fraction of CTTSs observed in the 
B30 and B35 clouds, located 2.2$^{\circ}$ and 2.7$^{\circ}$ from the central 
star $\lambda$~Orionis. They supposed that, as observed in the Trapezium around 
$\theta^{1}$ Ori \citep{Johnstone1998ApJ}, 
circumstellar disks could have been photo-evaporated by the far-UV radiation of 
the high-mass stars in the period before the supernova explosion, when high
and low-mass stars were confined by the parent cloud in a smaller region.
The same effect might have not affected the $\sigma$~Ori stars because its 
central high-mass star is less bright than $\lambda$~Orionis, and likely
also than
the supernova progenitor and because the $\sigma$~Ori cluster might never
have been confined in a smaller region as \cite{Dolan1999AJ, Dolan2001AJ}
supposed for the $\lambda$~Ori cluster. 

However, recent N-body simulations showed that the 
effect of photoevaporation of the circumstellar disks due to emission from 
the high-mass stars is negligible for clusters composed of less than 
1000 members \citep{Adams2006ApJ}, while \cite{Barrado2004ApJ} found only
170 candidate members
in their survey over an area of 0.3~deg$^2$ (with a completeness limit of
0.025~$M_{\odot}$), 
including most part of the $\lambda$~Ori cluster.
Moreover, \cite{Barrado2007ApJ} found several stars with disks near the
central O stars and no correlation between the fraction of stars with disks
and the distance from $\lambda$~Orionis.

Furthermore the difference in the accretion properties of stars with disks
in the two clusters, discussed in Sect.~\ref{par:acc_sori_lori}, does not
support the \cite{Dolan2001AJ} hypothesis, because the far-UV radiation of
high-mass stars could trigger the disk photoevaporation but should not
directly influence the accretion on the stars.
On the contrary, \cite{Fatuzzo2006ApJ} suggested that a supernova explosion
could increase the MARs of the nearby stars.
Specifically, they hypothesized that the enhancement of the cosmic rays flux
following a supernova explosion causes in the circumstellar disks of the
nearby stars an increase of the ionization levels and, therefore, of the
magneto-rotational instabilities, which, according to the
\cite{Gammie1996ApJ} model, trigger the accretion on the CTTSs

\subsection{The age hypothesis}

Another possibility is that the discrepancy between the two clusters could
be due to a different age. 
As shown by \cite{Hernandez2007ApJ} using \textit{Spitzer} photometry and by
\cite{Haisch2001ApJ} using $J$, $H$, $K$ and $L$ bands photometry, the
fraction of
stars with disks decreases from 100\% in the youngest regions to a few
percent in 6$-$7~Myr old clusters.
Moreover, \cite{Hernandez2007ApJ} and \cite{Barrado2007ApJ} found a larger
fraction of evolved thin disks in the older young clusters. Since a
significant fraction of these evolved disks might have stopped the accretion
onto the stars (\citealt{Sicilia2006ApJ} and references therein), it is
likely that in young clusters the fraction of non-accreting disks grows with
age.
Following these considerations, the suggestion that the $\lambda$~Ori
population is 
more evolved than the $\sigma$~Ori one could well explain the differences
between the 
two clusters.  

\begin{figure}
\resizebox{\hsize}{!}{
\includegraphics{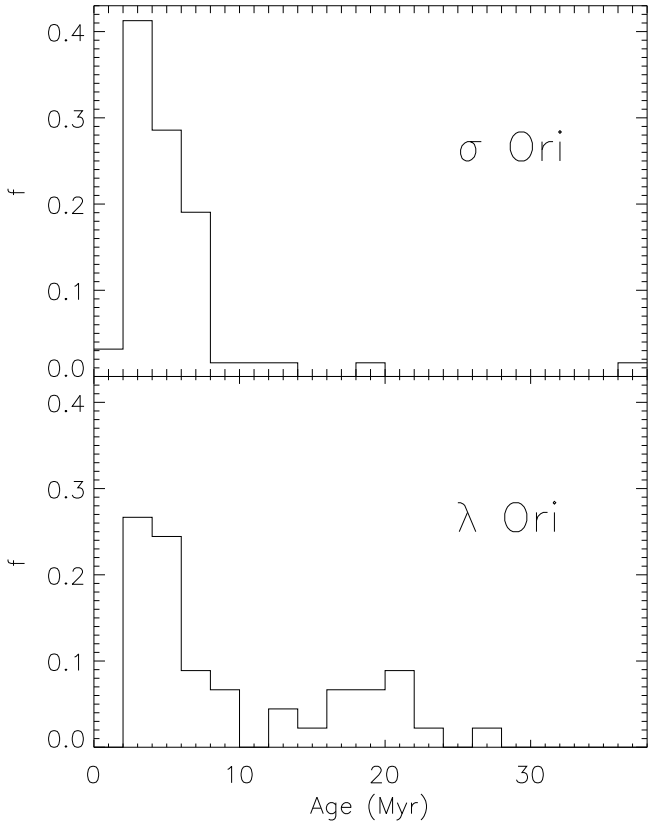}} 
\caption{Age distribution of the members selected in $\sigma$~Ori and
$\lambda$~Ori. Ages were derived from the $R$ and $I$ magnitudes using
\cite{Siess2000A&A} evolutionary models.
As discussed in the text, the age distributions shown in this figure are not
in agreement with the Li line pEWs shown in Fig.~\ref{fig:pew_col}.}
\label{fig:Age_histo}
\end{figure}  

To investigate this hypothesis, we plot in Fig.~\ref{fig:Age_histo} the age
distributions of the members selected in both clusters. Ages have been
calculated from the 
\cite{Siess2000A&A} evolutionary models using the $I$ and $R$ 
magnitudes, while for the distance modulus and the reddening we used the
same values 
as in Fig.~\ref{fig:targetCMD}.
The two histograms clearly show that the $\lambda$~Ori population is older
than the $\sigma$~Ori one. Moreover, the CMD for the two clusters, shown in
Fig.~\ref{fig:membCMD}, 
evidences that the age difference concerns especially the lowest mass stars.

Although Fig.~\ref{fig:Age_histo} seems to confirm the age hypothesis,
we note that the age distributions shown in Fig.~\ref{fig:Age_histo} could
be affected by 
large sources of errors: the reddening due to the presence of the
circumstellar disks could cause an underestimate of the age, while veiling
due to accretion could cause an overestimate of age; errors on distances,
which are poorly determined 
in both clusters, could strongly affect the age distributions and, as shown by 
\cite{Hillenbrand2007CS14}, different evolutionary models give different
results. The presence of some CTTSs among the oldest and lowest stars in $\lambda$~Ori
confirms that ages derived from R and I magnitudes are affected by these
sources of errors, in fact errors due to accretion veiling strongly affects 
the low-mass stars \citep{Herczeg2008ApJ}.
Moreover, taking into account that, according to the \cite{Siess2000A&A}
models, stars in the mass range 0.3$-$0.2~$M_{\odot}$ deplete very rapidly
their photospheric 
lithium after 15$-$30~Myr, the oldest stars of the $\lambda$~Ori cluster 
in this mass range should have depleted part of their lithium, but, as shown
in Fig.~\ref{fig:pew_col}, the lithium pEWs of $\lambda$~Ori members are
never below 450~m\AA.

More detailed HR diagrams based on spectral types derived from low-resolution spectroscopy are required to better investigate the age difference between the two clusters.

\section{Summary and conclusions}
\label{par:conclusions}

In this paper we have reported the results of FLAMES/VLT optical 
spectroscopic observations of 147 low-mass stars, in the
0.2$-$1.0~$M_{\sun}$ mass range, belonging to 
the two very similar clusters $\sigma$~Ori and $\lambda$~Ori. 
 
Using RVs, H$\alpha$ and Li line pEWs, we identified 65 bona-fide members 
of the $\sigma$~Ori cluster and 45 members of $\lambda$~Ori. Furthermore 
we discovered 16 new binary systems and binary candidates, 10
of which are probable members of the clusters and measured rotational velocities of
20 stars.
To study the accretion properties, we estimated the stellar 
MARs from the width at 10\% of the peak of the H$\alpha$ line
and  measured  the pEWs of the H$\alpha$ and other emission lines, which
are signatures of accretion/outflow phenomena.
We compared our results with the \textit{Spitzer} observations of the two
clusters
by \cite{Hernandez2007ApJ} and \cite{Barrado2007ApJ}, finding that:
a) the fraction of stars with disks obtained by \cite{Hernandez2007ApJ}
is likely underestimated due to the presence of a large number of field stars 
in their catalogue of members; b) the fraction of stars with a circumstellar
disk
in the $\sigma$~Ori cluster (52$\pm$9\%) is larger than in
$\lambda$~Ori (28$\pm$8\%); c)
the fraction of active disks in $\sigma$~Ori (78$\pm$16\%) is
larger than in $\lambda$~Ori (40$\pm$20\%).   
We have discussed two hypotheses that could explain the discrepancy between
the two clusters: either the circumstellar disks in the $\lambda$~Ori
cluster dissipate more rapidly due to the effect of the massive stars
emission or the $\lambda$~Ori cluster could be older and more evolved than
$\sigma$~Ori. 
The former hypothesis is in contradiction with some theoretical and
observational studies, while the latter one cannot be confirmed due to the
uncertainties in stellar ages. Further low-resolution spectroscopic
observations are required to reach a firm conclusion.

\begin{acknowledgements}

We thank the anonymous referee and Antonella Natta for useful suggestions and
comments. We acknowledge the staff of the ESO Data Management and Operations
department, who performed our observations in service mode.
 
This work makes use of results produced by the PI2S2 Project managed by the Consorzio
COMETA, a project co-funded by the Italian Ministry of University and
Research (MIUR) within the Piano Operativo Nazionale ``Ricerca Scientifica,
Sviluppo Tecnologico, Alta Formazione'' (PON 2000-2006). More information is
available at http://www.pi2s2.it and http://www.consorzio-cometa.it.

This research has been supported by an INAF grant on \textit{Stellar clusters as probes of star formation and early stellar evolution (PI: F. Palla).} 
This publication makes use of data products from the Two Micron All Sky Survey,
which is a joint project of the University of Massachusetts and the Infrared 
Processing and Analysis Center/California Institute of Technology, funded by 
the National Aeronautics and Space Administration and the National Science 
Foundation.
\end{acknowledgements}

\bibliographystyle{aa}
\bibliography{biblio}

\longtabL{1}{
\begin{landscape}
\small
\begin{longtable}{clcccccclrccllllll}
\caption{\label{tab:mag_star} Photometry and membership for the observed sample stars in the $\sigma$~Ori cluster.}\\
\hline \hline
ID& Name$^{\,a}$& RA$^{\,b}$\ \ \ \ \ \  DEC$^{\,b}$& $I^{\,c}$& $R$$-$$I^{\,c}$& ref$^{\,c}$& $J^{\,b}$& $H$$-$$K_\mathrm{s}^{\,b}$& SpT$^{\,d}$&
\multicolumn{1}{c}{$v_\mathrm{r}$}& \multicolumn{1}{c}{$v sin(i)$}& \multicolumn{1}{c}{Li pEW$^{\,e}$}& \multicolumn{4}{c}{Membership}& X$^f$& Notes\\
&  & \multicolumn{1}{c}{(J2000)}& & & & & & & \multicolumn{1}{c}{(km/s)}& \multicolumn{1}{c}{(km/s)}& \multicolumn{1}{c}{(m\AA)}& $v_\mathrm{r}$& Li& H$\alpha$& \multicolumn{1}{c}{Tot}& & \\
\hline
\endfirsthead
\caption{continued}\\
\hline \hline
ID& Name$^{\,a}$& RA$^{\,b}$ \ \ \ \ \ \  DEC$^{\,b}$& $I^{\,c}$& $R$$-$$I^{\,c}$& ref$^{\,c}$& $J^{\,b}$& $H$$-$$K_\mathrm{s}^{\,b}$& SpT$^{\,d}$&
  \multicolumn{1}{c}{$v_\mathrm{r}$}& \multicolumn{1}{c}{$v sin(i)$}& \multicolumn{1}{c}{Li pEW$^{\,e}$}& \multicolumn{4}{c}{Membership}& X$^f$& Notes \\
   &  & \multicolumn{1}{c}{(J2000)}& & & & & & & \multicolumn{1}{c}{(km/s)}& \multicolumn{1}{c}{(km/s)}& \multicolumn{1}{c}{(m\AA)}& $v_\mathrm{r}$& Li& H$\alpha$& \multicolumn{1}{c}{Tot}& & \\
\hline
\endhead
\hline
\endfoot
\hline
\multicolumn{17}{l}{$a$: star names are from the following sources:
4771-..., r05..., p05... $=$ \citet{Wolk1996PhDT}; SE $=$ \citet{Scholz2004A&A}; SWW $=$\citet{Sherry2004AJ}; J05... $=$ }\\
\multicolumn{17}{l}{\citet{Zapatero2002A&A,Bejar2004AN}: note that we have dropped the S\,Ori prefix in front of the J05... names; K $=$
\citet{Kenyon2005MNRAS}; }\\
\multicolumn{17}{l}{B $=$ \citet{Burningham2005MNRAS}; 2M $=$ 2MASS catalogue.} \\
\multicolumn{17}{l}{$b$: coordinates and infrared photometry are from the 2MASS catalogue.}\\
\multicolumn{17}{l}{$c$: optical photometric data are taken from the following sources:
(1) \cite{Sherry2004AJ}, (2) \cite{Wolk1996PhDT}, (3) \cite{Kenyon2005MNRAS}.} \\
\multicolumn{17}{l}{$d$: spectral types marked with asterisk have been derived spectroscopically by
\cite{Zapatero2002A&A}. The other spectral types have been derived from}\\
\multicolumn{17}{l}{photometry (see Sect.~\ref{par:T_selection}).} \\
\multicolumn{17}{l}{$e$: for double-lined spectroscopic binaries, we only indicate whether the line is identified (Y) or not (N)}\\
\multicolumn{17}{l}{$f$: Y = detected, N = undetected, - = outside the XMM field of view}
\endlastfoot
S01& p053834-0239       & 5 38 34.79 \ $-$2 39 30.0& 11.91& 1.04& 2& 10.44& 0.25& M1.0    &  46.21$\pm$0.64      &             & 72$\pm$6 & N & N & N & N & N &               \\
S02& r053838-0236       & 5 38 38.22 \ $-$2 36 38.4& 12.38& 0.86& 2& 11.16& 0.15& K8.0$^*$&  30.93$\pm$0.33      & $<$17.8     & 551$\pm$12& Y & Y & Y & Y & Y &               \\
S03& r053835-0231       & 5 38 35.47 \ $-$2 31 51.7& 12.45& 0.79& 2& 11.30& 0.17& K7.5    &  30.60$\pm$0.29      & $<$17.0     & 582$\pm$4 & Y & Y & Y & Y & Y &               \\
S04& 4771-1097          & 5 38 35.87 \ $-$2 30 43.3& 12.47& 0.81& 2& 11.24& 0.17& K8.0$^*$& \multicolumn{1}{c}{?}&             & 338$\pm$32& ? & Y & Y & Y?& Y & SB            \\
S05& 4771-0910          & 5 39 18.83 \ $-$2 30 53.1& 12.58& 0.84& 2& 11.40& 0.30& K9.0   &  29.14$\pm$0.65       & 42.7$^{+1.6}_{-3.1}$ & 524$\pm$39& Y & Y & Y & Y & Y &               \\
S06& 4771-1092          & 5 39 07.61 \ $-$2 32 39.2& 12.63& 0.81& 2& 11.30& 0.31& K8.0    &  30.27$\pm$0.32      & $<$17.0     & 532$\pm$8 & Y & Y & Y & Y & Y &               \\
S07& 4771-1075          & 5 39 05.41 \ $-$2 32 30.3& 12.66& 0.87& 2& 11.55& 0.19& K7.0$^*$&  30.40$\pm$0.20      & $<$17.4     & 593$\pm$12& Y & Y & Y & Y & Y &               \\
S08& 4771-0579          & 5 39 30.43 \ $-$2 35 07.3& 12.71& 0.74& 2& 11.61& 0.15& K6.5    &  80.50$\pm$0.80      &              &30$\pm$5 & N & N & N & N & N &               \\
S09& p053933-0236       & 5 39 33.44 \ $-$2 36 41.9& 12.73& 0.64& 2& 11.85& 0.14& K5.0    & $-$14.01$\pm$1.15    &              & 29$\pm$3 & N & N & N & N & N &               \\
S10& r053841-0237       & 5 38 41.29 \ $-$2 37 22.6& 12.77& 0.87& 2& 11.46& 0.21& K9.5    &  30.89$\pm$0.30      & $<$17.0     & 592$\pm$9 & Y & Y & Y & Y & Y &               \\
S11& SWW166             & 5 38 53.07 \ $-$2 38 53.6& 12.78& 1.07& 1& 11.62& 0.21& M1.0    & $-$0.10$\pm$0.52     &             & 35$\pm$7 & N & N & N & N & Y &               \\
S12& r053840-0230       & 5 38 40.27 \ $-$2 30 18.5& 12.80& 0.94& 2& 11.51& 0.37& M0.0$^*$&  31.51$\pm$0.36      & $<$17.0     & 552$\pm$19& Y & Y & Y & Y & Y &               \\
S13& 4771-0041          & 5 38 27.26 \ $-$2 45 09.7& 12.82& 0.82& 2& 11.95& 0.85& K7.0$^*$&  30.00$\pm$0.49      & $<$17.0     & 431$\pm$11& Y & Y & Y & Y & Y &               \\
S14& SWW135             & 5 39 25.20 \ $-$2 38 22.0& 12.83& 0.95& 1& 11.31& 0.45& M0.5    &  30.10$\pm$0.46      & $<$17.0     & 480$\pm$7 & Y & Y & Y & Y & Y &               \\
S15& SWW78              & 5 38 47.46 \ $-$2 35 25.2& 12.86& 0.99& 1&    - & 0.25& M1.0    &  30.32$\pm$0.40      & $<$17.0     & 589$\pm$6 & Y & Y & Y & Y & Y &               \\
S16& r053832-0235b      & 5 38 32.84 \ $-$2 35 39.2& 12.88& 0.83& 2& 11.54& 0.17& K8.5    &  30.12$\pm$0.27      & $<$17.0     & 576$\pm$6 & Y & Y & Y & Y & Y &               \\
S17& 4771-0080          & 5 38 52.01 \ $-$2 46 43.7& 12.91& 0.85& 3& 11.52& 0.35& K9.0    &  30.51$\pm$0.36      & $<$17.0     & 551$\pm$10& Y & Y & Y & Y & Y &               \\
S18& 4771-1038          & 5 39 11.63 \ $-$2 36 02.9& 12.93& 0.95& 1& 11.62& 0.22& K8.0$^*$&  30.13$\pm$0.28      & $<$17.0     & 590$\pm$6 & Y & Y & Y & Y & Y &               \\
S19& r053849-0238       & 5 38 49.17 \ $-$2 38 22.2& 12.96& 1.03& 1& 11.39& 0.15& M0.5$^*$&  30.79$\pm$0.28      & $<$17.0     & 623$\pm$5 & Y & Y & Y & Y & Y &               \\
S20& 4771-0961          & 5 38 02.21 \ $-$2 29 55.6& 12.99& 0.69& 2& 11.83& 0.19& K6.0    &  21.24$\pm$0.69      &             & 32$\pm$4 & N & N & N & N & N &               \\
S21& SWW102             & 5 38 47.92 \ $-$2 37 19.2& 13.02& 1.18& 1&    - &   - & M1.0    &  31.77$\pm$0.51      & $<$17.0     & 481$\pm$8 & Y & Y & Y & Y & Y &               \\
S22& r053830-0241       & 5 38 30.60 \ $-$2 41 18.8& 13.06& 0.47& 2& 12.42& 0.12& K2.0    &  29.17$\pm$0.83      &             &  22$\pm$5 & Y & N & N & N & N &               \\
S23& SWW48              & 5 38 42.28 \ $-$2 37 14.8& 13.06& 0.98& 1& 11.77& 0.22& M0.5    &  30.42$\pm$0.43      & $<$17.0     & 568$\pm$39& Y & Y & Y & Y & N &               \\
S24& SWW36              & 5 38 43.55 \ $-$2 33 25.4& 13.09& 1.06& 1& 11.72& 0.24& M1.0    &  31.63$\pm$0.33      & $<$17.0     & 620$\pm$23& Y & Y & Y & Y & Y &               \\
S25& 4771-1090          & 5 38 46.05 \ $-$2 43 47.8& 13.12& 0.74& 2& 11.97& 0.16& K6.5    &   8.38$\pm$0.60      &             & 39$\pm$4 & N & N & N & N & N &               \\
S26& SWW205             & 5 38 49.22 \ $-$2 41 25.1& 13.23& 1.50& 1& 11.67& 0.27& M3.5    & \multicolumn{1}{c}{?}&            & 442$\pm$57& ? & Y & Y & Y?& Y & SB            \\
S27& r053838-0226       & 5 38 38.56 \ $-$2 26 44.8& 13.30& 0.71& 3& 12.30& 0.16& K6.0    &  91.40$\pm$0.68      &            &  33$\pm$2 & N & N & N & N & N &               \\
S28& SWW35              & 5 38 48.68 \ $-$2 36 16.2& 13.37& 0.87& 1& 12.11& 0.20& K9.5    &  32.53$\pm$0.46      & $<$17.0     & 590$\pm$4 & Y & Y & Y & Y & Y &               \\
S29& r053851-0236       & 5 38 51.45 \ $-$2 36 20.6& 13.38& 0.87& 2& 12.44& 0.26& K9.5    &  32.06$\pm$0.73      &            & \multicolumn{1}{c}{Y}& Y & Y & Y & Y & Y & SB2\\
S30& K3.01-170          & 5 38 15.90 \ $-$2 34 41.2& 13.42& 0.64& 3& 12.37& 0.15& K5.0    & $-$30.56$\pm$0.70    &            &  47$\pm$2 & N & N & N & N & N &               \\
S31& r053812-0232       & 5 38 12.60 \ $-$2 33 01.5& 13.44& 0.62& 3& 12.46& 0.16& K5.0    &  33.15$\pm$0.71      &            &  21$\pm$4 & Y & N & N & N & N &               \\
S32& SWW195             & 5 39 11.52 \ $-$2 31 06.6& 13.45& 0.95& 1& 11.99& 0.46& M0.5    &  30.91$\pm$0.79      & $<$17.0     & 263$\pm$4 & Y & Y & Y & Y & Y &               \\
S33& SWW97              & 5 38 45.38 \ $-$2 41 59.4& 13.46& 1.16& 1& 11.99& 0.29& M1.0    &  31.86$\pm$0.57      & $<$17.0     & 482$\pm$10& Y & Y & Y & Y & Y &               \\
S34& r053831-0235       & 5 38 31.58 \ $-$2 35 14.9& 13.49& 1.11& 1& 11.52& 0.35& M0.0$^*$&  32.13$\pm$0.55      & $<$19.2     & 479$\pm$6 & Y & Y & Y & Y & Y &               \\
S35& SWW47              & 5 38 53.17 \ $-$2 43 52.8& 13.49& 1.08& 1& 12.23& 0.20& M1.0    &  30.48$\pm$0.49      & $<$17.0     & 561$\pm$8 & Y & Y & Y & Y & Y &               \\
S36& K4.03-511          & 5 38 19.15 \ $-$2 35 27.9& 13.51& 0.74& 3& 12.31& 0.18& K6.5    & 121.73$\pm$0.67      &             &  59$\pm$17& N & N & N & N & N &               \\
S37& J053920.5-022737   & 5 39 20.44 \ $-$2 27 36.8& 13.52& 1.29& 1& 12.15& 0.26& M2.0$^*$&  23.73$\pm$0.45      &             & 581$\pm$7 & N & Y & Y & N?& Y &               \\
S38& SWW71              & 5 38 55.44 \ $-$2 41 29.7& 13.59& 1.39& 1& 12.17& 0.28& M3.0    &   3.53$\pm$0.54      &             &  78$\pm$5 & N & N & N & N & N &               \\
S39& SWW87              & 5 38 27.74 \ $-$2 43 01.0& 13.67& 1.33& 1& 12.19& 0.16& M2.5    &  29.91$\pm$0.58      & 21.9$^{+0.5}_{-3.1}$ & 549$\pm$32& Y & Y & Y & Y & Y &               \\
S40& p053902-0238       & 5 39 02.16 \ $-$2 38 38.2& 13.69& 1.67& 2& 12.40& 0.16& M4.0    &  61.34$\pm$0.71      &             & 33$\pm$4 & N & N & N & N & N &               \\
S41& r053833-0236       & 5 38 34.06 \ $-$2 36 37.5& 13.72& 1.52& 1& 11.98& 0.25& M3.5$^*$&  34.12$\pm$2.21      & 45.7$^{+1.1}_{-5.3}$ & 424$\pm$56& Y & Y & Y & Y & Y &               \\
S42& p053925-0231       & 5 39 25.34 \ $-$2 31 43.7& 13.75& 0.81& 2& 12.43& 0.26& K8.0    & 103.44$\pm$0.61      &            &  65$\pm$6 & N & N & N & N & N &               \\
S43& SWW41              & 5 38 08.27 \ $-$2 35 56.3& 13.76& 1.32& 1& 12.14& 0.33& M2.5    & \multicolumn{1}{c}{?}&            & 518$\pm$4 & ? & Y & Y & Y?& Y & SB            \\
S44& SWW50              & 5 38 31.41 \ $-$2 36 33.8& 13.80& 1.29& 1& 12.17& 0.49& M2.0    &  30.81$\pm$0.88      & $<$17.0     & 342$\pm$2 & Y & Y & Y & Y & N &               \\
S45& K1.02-7            & 5 38 48.29 \ $-$2 36 41.0& 13.83& 1.74& 3& 12.04& 0.26& M4.5    &  30.21$\pm$1.38      & 32.1$^{+12.4}_{-7.1}$ & 510$\pm$60& Y & Y & Y & Y & Y &               \\
S46& r053832-0235a      & 5 38 32.60 \ $-$2 35 04.1& 13.86& 0.47& 3& 13.27& 0.17& K2.0    &  51.92$\pm$1.49      &            &  23$\pm$6 & N & N & N & N & N &               \\
S47& r053834-0234       & 5 38 34.31 \ $-$2 35 00.1& 13.88& 0.43& 2&    - &   - & K1.0    &  30.37$\pm$0.38      & 29.3$^{+2.8}_{-5.6}$ & 542$\pm$34& Y & Y & Y & Y & Y &               \\
S48& r053814-0235       & 5 38 14.23 \ $-$2 35 07.3& 13.99& 0.59& 3& 13.11& 0.11& K4.5    &  48.50$\pm$0.71      &            & 42$\pm$13 & N & N & N & N & N &               \\
S49& SWW177             & 5 38 29.12 \ $-$2 36 02.7& 14.04& 1.18& 1&    - &   - & M1.0    &  32.46$\pm$0.43      & $<$17.0    & 604$\pm$7 & Y & Y & Y & Y & Y &               \\
S50& SWW88              & 5 38 05.67 \ $-$2 40 19.4& 14.07& 1.27& 1& 12.77& 0.28& M2.0    & \multicolumn{1}{c}{?}&            &  94$\pm$5 & ? & N & N & N & N & SB            \\
S51& SE34               & 5 38 50.39 \ $-$2 26 47.7& 14.10& 1.48& 1& 12.50& 0.29& M3.5    &  30.48$\pm$0.57      & 21.5$^{+6.3}_{-4.5}$ & 582$\pm$23& Y & Y & Y & Y & Y &               \\
S52& SE3                & 5 38 13.20 \ $-$2 26 08.8& 14.10& 1.57& 1& 12.48& 0.27& M4.0    &  32.07$\pm$0.52      & $<$17.0     & 633$\pm$14& Y & Y & Y & Y & Y &               \\
S53& SWW144             & 5 38 43.34 \ $-$2 32 00.8& 14.12& 1.75& 1& 12.24& 0.28& M4.5    &  30.61$\pm$1.06      &             & \multicolumn{1}{c}{Y}& Y & Y & Y & Y & Y & SB2\\
S54& SWW101             & 5 38 49.64 \ $-$2 45 26.9& 14.15& 1.23& 1& 13.15& 0.19& M1.5    &  41.57$\pm$0.57      &             &  88$\pm$7 & N & N & N & N & - &               \\
S55& SE51               & 5 39 17.17 \ $-$2 25 43.4& 14.18& 1.15& 1& 12.90& 0.19& M1.0    &  29.47$\pm$0.48      & $<$17.0     & 133$\pm$8 & Y & N & Y & Y & Y &               \\
S56& r053923-0233       & 5 39 22.87 \ $-$2 33 33.1& 14.19& 1.21& 1& 12.83& 0.26& M2.0$^*$&  30.26$\pm$0.37      & $<$17.0     & 584$\pm$7 & Y & Y & Y & Y & Y &               \\
S57& SWW200             & 5 38 49.93 \ $-$2 41 22.8& 14.21& 1.37& 1& 12.75& 0.22& M3.0    &  31.35$\pm$0.49      & $<$17.0     & 604$\pm$5 & Y & Y & Y & Y & Y &               \\
S58& SWW28              & 5 39 02.77 \ $-$2 29 55.8& 14.21& 1.51& 1& 12.61& 0.31& M3.5    &  31.77$\pm$0.49      & $<$17.0     & 603$\pm$9 & Y & Y & Y & Y & Y &               \\
S59& SWW127             & 5 39 24.36 \ $-$2 34 01.4& 14.28& 1.25& 1& 12.98& 0.21& M2.0    &  33.18$\pm$0.45      & $<$17.0     &  47$\pm$21& Y & N & Y & Y & Y &               \\
S60& J053836.7-024414   & 5 38 36.69 \ $-$2 44 13.7& 14.32& 1.76& 1& 12.54& 0.27& M4.5    &  30.57$\pm$0.75      & $<$17.0     & 570$\pm$3 & Y & Y & Y & Y & N &               \\
S61& 4771-1099          & 5 39 07.81 \ $-$2 40 09.1& 14.32& 0.60& 2& 13.57& 0.10& K4.5    &  45.54$\pm$1.07      &             & 42$\pm$16& N & N & N & N & N &               \\
S62& J053820.1-023802   & 5 38 20.22 \ $-$2 38 01.6& 14.33& 1.69& 1& 12.58& 0.25& M4.0$^*$&  33.92$\pm$2.54      & 56.2$^{+7.0}_{-9.8}$   & 480$\pm$36& Y & Y & Y & Y & Y &               \\
S63& J053827.5-023504   & 5 38 27.51 \ $-$2 35 04.2& 14.35& 1.36& 3& 12.83& 0.25& M3.5$^*$&  29.72$\pm$0.63      & $<$17.0     & 516$\pm$8 & Y & Y & Y & Y & Y &               \\
S64& r053907-0228       & 5 39 07.59 \ $-$2 28 23.4& 14.37& 1.45& 1& 12.88& 0.19& M3.0$^*$&  31.48$\pm$0.46      & $<$17.0     & 615$\pm$11& Y & Y & Y & Y & Y &               \\
S65& SWW25              & 5 38 41.60 \ $-$2 30 28.9& 14.39& 1.45& 1& 12.84& 0.21& M3.5    &  31.01$\pm$0.51      & $<$17.0     & 588$\pm$14& Y & Y & Y & Y & N &               \\
S66& K1.01-114          & 5 39 36.06 \ $-$2 36 31.0& 14.43& 0.76& 3& 13.07& 0.22& K7.0    &  48.55$\pm$0.63      &             &  56$\pm$23& N & N & N & N & N &               \\
S67& p053841-0236       & 5 38 41.36 \ $-$2 36 44.5& 14.49& 1.46& 3& 12.99& 0.22& M3.5    &  31.92$\pm$0.51      & $<$17.0     & 577$\pm$14& Y & Y & Y & Y & Y &               \\
S68& SWW86              & 5 38 40.54 \ $-$2 33 27.6& 14.54& 1.65& 1& 12.80& 0.27& M4.0    &  31.79$\pm$0.73      & $<$17.0     &  522$\pm$11& Y & Y & Y & Y & N &               \\
S69& SWW31              & 5 38 39.03 \ $-$2 45 32.2& 14.58& 1.34& 1& 12.91& 0.30& M2.5    &  30.67$\pm$0.56      & $<$17.0     & 551$\pm$10& Y & Y & Y & Y & N &               \\
S70& K1.02-91           & 5 38 56.23 \ $-$2 31 15.4& 14.58& 0.76& 3& 13.42& 0.25& K7.0    &  69.74$\pm$0.63      &             & 37$\pm$9 & N & N & N & N & N &               \\
S71& SWW29              & 5 38 47.19 \ $-$2 34 36.8& 14.64& 1.45& 1& 12.56& 0.48& M3.5    &  31.87$\pm$1.15      & $<$17.0     & 306$\pm$9 & Y & Y & Y & Y & N &               \\
S72& J053834.5-024109   & 5 38 34.60 \ $-$2 41 08.8& 14.72& 1.46& 1& 13.10& 0.33& M3.5    &  30.88$\pm$0.59      & $<$17.0     & 547$\pm$3 & Y & Y & Y & Y & N &               \\
S73& SWW15              & 5 38 43.87 \ $-$2 37 06.8& 14.80& 1.79& 1& 12.84& 0.37& M4.5    &  31.23$\pm$0.69      & $<$17.0     & 453$\pm$7 & Y & Y & Y & Y & N &               \\
S74& SWW227             & 5 38 59.23 \ $-$2 33 51.4& 14.83& 1.26& 1& 12.89& 0.58& M2.0    &  31.58$\pm$0.70      & $<$17.0     & 395$\pm$33& Y & Y & Y & Y & Y &               \\
S75& J053914.5-022834   & 5 39 14.47 \ $-$2 28 33.3& 14.85& 1.52& 1& 13.34& 0.31& M3.5$^*$&  29.19$\pm$0.53      & $<$17.0     & 61$\pm$6 & Y & N & Y & Y & Y &               \\
S76& K1.02-4            & 5 39 15.83 \ $-$2 36 50.7& 14.86& 1.50& 3& 13.25& 0.32& M3.5    &  31.27$\pm$0.60      & $<$17.0     & 579$\pm$14& Y & Y & Y & Y & Y &               \\
S77& SWW230             & 5 38 42.86 \ $-$2 38 52.5& 14.92& 1.11& 1& 13.68& 0.27& M1.0    &  41.98$\pm$0.50      &             & 70$\pm$6 & N & N & N & N & N &               \\
S78& SWW129             & 5 39 08.78 \ $-$2 31 11.5& 14.97& 1.63& 1& 13.04& 0.45& M4.0    &  31.91$\pm$0.64      & $<$17.0     & 470$\pm$13& Y & Y & Y & Y & N &               \\
S79& K1.02-19           & 5 38 51.74 \ $-$2 36 03.3& 14.97& 1.87& 3& 12.91& 0.26& M5.0    &  30.81$\pm$0.76      & $<$17.0     & 524$\pm$10& Y & Y & Y & Y & Y &               \\
S80& K3.01-167          & 5 38 49.70 \ $-$2 34 52.6& 15.03& 1.81& 3& 12.98& 0.25& M4.5    &  31.15$\pm$0.80      & $<$17.0     & 483$\pm$4 & Y & Y & Y & Y & N &               \\
S81& K8                 & 5 38 50.78 \ $-$2 36 26.8& 15.06& 2.00& 3& 13.11& 0.24& M5.0    &  31.09$\pm$1.51      & 24.5$^{+6.9}_{-2.1}$ & 442$\pm$37& Y & Y & Y & Y & N &               \\
S82& K12                & 5 39 13.48 \ $-$2 23 51.9& 15.18& 1.41& 3& 13.95& 0.24& M3.0    &  40.63$\pm$0.63      &              & 135$\pm$57& N & N & N & N & - &               \\
S83& B3.01-67           & 5 38 46.85 \ $-$2 36 43.5& 15.28& 1.88& 3& 13.22& 0.33& M5.0    &  30.22$\pm$1.39      & 18.8$^{+2.7}_{-1.4}$ & 502$\pm$3 & Y & Y & Y & Y & N &               \\
S84& SWW222             & 5 39 30.56 \ $-$2 38 27.0& 15.30& 1.32& 3& 13.81& 0.22& M2.5    &  23.43$\pm$0.69      &              & \multicolumn{1}{c}{N}& N & N & Y & N & Y & SB2\\
S85& K1.02-156          & 5 38 51.01 \ $-$2 27 45.7& 15.51& 1.00& 3& 14.28& 0.23& M1.0    &  72.34$\pm$0.88      &              & \multicolumn{1}{c}{N}& N & N & N & N & Y & SB2\\
S86& K3.01-263          & 5 38 28.25 \ $-$2 32 27.4& 15.62& 0.53& 3& 14.86& 0.15& K3.5    & 113.37$\pm$1.07      &              &  43$\pm$13& N & N & N & N & Y &               \\
S87& K4.03-422          & 5 38 45.28 \ $-$2 37 29.3& 15.67& 1.79& 3& 13.37& 0.51& M4.5    &  30.19$\pm$0.91      & $<$17.0     & 402$\pm$3 & Y & Y & Y & Y & N &               \\
S88& K3.01-226          & 5 38 35.29 \ $-$2 33 13.1& 15.96& 1.62& 3& 13.91& 0.32& M4.0    &  31.03$\pm$0.98      &             & 582$\pm$7 & Y & Y & Y & Y & N &               \\
S89& r053808-0235b      & 5 38 08.66 \ $-$2 35 41.4& 15.99& 0.47& 3& 15.23& 0.14& K2.0    &  16.54$\pm$2.02      &             & \multicolumn{1}{c}{N}& N & N & N & N & N & SB2\\
S90& p053840-0232       & 5 38 40.89 \ $-$2 32 59.8& 16.00& 0.50& 3& 15.36& 0.14& K3.0    &  62.12$\pm$1.50      &             &  64$\pm$6 & N & N & N & N & N &               \\
S91& p053902-0222       & 5 39 02.94 \ $-$2 22 41.8& 16.02& 1.36& 3& 14.79& 0.17& M3.0    & \multicolumn{1}{c}{?}&             & 139$\pm$8 & ? & N & N & N & N & SB            \\
S92& J053826.8-022846   & 5 38 26.84 \ $-$2 38 46.0& 16.12& 1.91& 3& 14.11& 0.28& M5.0    &  30.66$\pm$0.94      & $<$17.0     & 494$\pm$9 & Y & Y & Y & Y & N &               \\
S93& K1.02-18           & 5 38 42.39 \ $-$2 36 04.4& 16.26& 1.49& 3& 14.26& 0.38& M3.5    &  29.66$\pm$1.26      & $<$17.0     & 272$\pm$20& Y & Y & Y & Y & N &               \\
S94& K4.03-1807         & 5 38 41.46 \ $-$2 35 52.3& 16.32& 2.25& 3& 14.00& 0.37& M6.0    &  30.40$\pm$1.48      & $<$17.0     & 322$\pm$27& Y & Y & Y & Y & N &               \\
S95& 2MJ05391483-0244415& 5 39 14.83 \ $-$2 44 41.6&   -- &  -- & -& 12.34& 0.19& --      &  61.23$\pm$0.59      &             &   32$\pm$3 & N & N & N & N & N &               \\
S96& K3.01-325          & 5 38 32.68 \ $-$2 31 15.6&   -- &  -- & -& 12.55& 0.08& --      & $-$34.79$\pm$1.19    &             &   9$\pm$4 & N & N & N & N & Y & SB1           \\
S97& 2MJ05391003-0242425& 5 39 10.04 \ $-$2 42 42.5&   -- &  -- & -& 12.97& 0.24& --      &  78.96$\pm$0.68      &             & 70$\pm$6 & N & N & N & N & N &               \\
S98& 2MJ05390458-0241493& 5 39 04.59 \ $-$2 41 49.4&   -- &  -- & -& 13.96& 0.69& --      &  31.61$\pm$0.73      & $<$17.0     & 443$\pm$57& Y & Y & Y & Y & N &               \\
\end{longtable}
\end{landscape}
}

\longtabL{2}{
\begin{landscape}
\small
\begin{longtable}{clcccccclrcclllll}
\caption{\label{tab:star_lambda} Photometry and membership for the observed sample stars in the $\lambda$~Ori cluster.}\\
\hline \hline
ID& Name$^{\,a}$& RA$^{\,b}$\ \ \ \ \ \  DEC$^{\,b}$& $I^{\,c}$& $R$$-$$I^{\,c}$& ref$^{\,c}$& $J^{\,b}$& $H$$-$$K_\mathrm{s}^{\,b}$& SpT$^{\,d}$&
\multicolumn{1}{c}{$v_\mathrm{r}$}& \multicolumn{1}{c}{$v sin(i)$}& \multicolumn{1}{c}{Li pEW$^{\,e}$}& \multicolumn{4}{c}{Membership}& Notes\\
&  & \multicolumn{1}{c}{(J2000)}& & & & & & & \multicolumn{1}{c}{(km/s)}& \multicolumn{1}{c}{(km/s)}& \multicolumn{1}{c}{(m\AA)}& $v_\mathrm{r}$& Li& H$\alpha$& \multicolumn{1}{c}{Tot}& \\
\hline
\endfirsthead
\caption{continued.}\\
\hline \hline
ID& Name$^{\,a}$& RA$^{\,b}$\ \ \ \ \ \  DEC$^{\,b}$& $I^{\,c}$& $R$$-$$I^{\,c}$& ref$^{\,c}$& $J^{\,b}$& $H$$-$$K_\mathrm{s}^{\,b}$& SpT$^{\,d}$&
\multicolumn{1}{c}{$v_\mathrm{r}$}& \multicolumn{1}{c}{$v sin(i)$} & \multicolumn{1}{c}{Li pEW$^{\,e}$}& \multicolumn{4}{c}{Membership}& Notes\\
&  & \multicolumn{1}{c}{(J2000)}& & & & & & & \multicolumn{1}{c}{(km/s)}& \multicolumn{1}{c}{(km/s)}& \multicolumn{1}{c}{(m\AA)}& $v_\mathrm{r}$& Li& H$\alpha$& \multicolumn{1}{c}{Tot}& \\
\hline
\endhead
\hline
\endfoot
\hline
\multicolumn{16}{l}{$a$: star names are from the following sources: DM $=$ \cite{Dolan1999AJ}; LOri $=$ \cite{Barrado2004ApJ}.}\\
\multicolumn{16}{l}{$b$: coordinates and infrared photometry are from the 2MASS catalogue.}\\
\multicolumn{16}{l}{$c$: optical photometric data are taken from: (1) \cite{Dolan1999AJ}, (2) \cite{Barrado2004ApJ}.}\\
\multicolumn{16}{l}{$d$: spectral types marked with asterisk have been derived spectroscopically by \cite{Barrado2004ApJ}.
The other spectral types have been derived from}\\
\multicolumn{16}{l}{photometry (see Sect.~\ref{par:T_selection}).} \\
\multicolumn{16}{l}{$e$: for the double-lined spectroscopic binary L10 and for the star L48, we only indicate whether the line is identified (Y) or not (N)}
\endlastfoot
L01& LOri-003& 5 35 55.43 \  9 56 30.9& 12.65& 0.74& 2& 11.42& 0.21& K8.5    & \multicolumn{1}{c}{?}&                      & 520$\pm$32 & ? & Y & Y & Y?& SB\\
L02& DM009   & 5 34 32.83 \  9 59 30.9& 12.94& 0.78& 1& 11.84& 0.17& K4.5    &  27.09$\pm$0.26      & 18.9$^{+2.7}_{-1.9}$ & 464$\pm$23 & Y & Y & Y & Y &   \\
L03& LOri-016& 5 35 13.46 \  9 55 25.3& 13.18& 0.89& 2& 11.96& 0.23& K8.0    & \multicolumn{1}{c}{?}&                      & 531$\pm$33 & ? & Y & Y & Y?& SB\\    
L04& DM014   & 5 34 39.21 \  9 52 55.3& 13.30& 1.06& 1& 12.07& 0.17& K7.0    &  27.72$\pm$0.18      &                           $<$17.0         &  523$\pm$26 & Y & Y & Y & Y &   \\   
L05& LOri-020& 5 34 57.57 \  9 46 07.2& 13.31& 1.34& 2& 11.86& 0.19& M2.0    &  21.43$\pm$0.53      &                      & 18$\pm$29 & N & N & Y & N &   \\   
L06& LOri-022& 5 35 51.34 \  9 55 11.2& 13.38& 1.03& 2& 12.10& 0.25& M1.5    &  28.40$\pm$0.38      &      $<$17.0         & 516$\pm$33 & Y & Y & Y & Y &   \\   
L07& LOri-024& 5 34 57.12 \  9 54 36.7& 13.45& 0.98& 2& 12.14& 0.23& M0.0    &  27.28$\pm$0.29      &      $<$17.0         & 554$\pm$21 & Y & Y & Y & Y &   \\   
L08& LOri-026& 5 34 36.23 \  9 53 44.2& 13.47& 1.10& 2& 12.05& 0.23& K9.5    &  27.29$\pm$0.43      & 26.0$^{+6.0}_{-1.0}$ & 581$\pm$43 & Y & Y & Y & Y &   \\   
L09& DM019   & 5 34 48.46 \  9 57 15.7& 13.73& 1.20& 1& 12.42& 0.18& M0.5    &  27.47$\pm$0.28      &      $<$17.0         & 571$\pm$27 & Y & Y & Y & Y &   \\   
L10& LOri-030& 5 35 12.54 \  9 55 19.5& 13.74& 1.21& 2& 12.43& 0.26& M1.5    & \multicolumn{1}{c}{?}&                      & \multicolumn{1}{c}{Y}& ? & Y & Y & Y? & SB2\\    
L11& LOri-034& 5 35 19.92 \ 10 02 36.5& 13.97& 1.13& 2& 12.44& 0.46& M1.0    &  24.36$\pm$0.30      &      $<$17.0         & 541$\pm$35 & N & Y & Y & Y?&   \\     
L12& LOri-035& 5 35 15.14 \ 10 01 06.8& 13.97& 1.28& 2& 12.55& 0.23& M2.5    &  21.36$\pm$0.72      & 30.2$^{+10.0}_{-6.2}$& 572$\pm$21 & N & Y & Y & Y?&   \\   
L13& LOri-036& 5 34 39.29 \ 10 01 28.7& 13.98& 1.49& 2& 12.58& 0.23& M3.0    &  24.63$\pm$0.38      &                      & 56$\pm$22 & N & N & Y & N &   \\   
L14& LOri-037& 5 34 35.61 \  9 59 43.3& 13.99& 1.18& 2& 12.46& 0.24& M2.0    &  27.13$\pm$0.33      &      $<$17.0         & 566$\pm$57 & Y & Y & Y & Y &   \\
L15& LOri-040& 5 35 39.48 \  9 50 32.8& 14.06& 1.32& 2& 12.55& 0.29& M3.0    &  26.95$\pm$0.46      &      $<$21.3         & 561$\pm$45 & Y & Y & Y & Y &   \\   
L16& LOri-041& 5 35 30.45 \  9 50 34.1& 14.10& 1.45& 2& 12.50& 0.27& M3.5    &  27.05$\pm$0.71      & 61.9$^{+11.8}_{-5.7}$& 485$\pm$40 & Y & Y & Y & Y &   \\   
L17& DM030   & 5 35 17.17 \  9 51 11.4& 14.13& 1.46& 1& 12.80& 0.22& M1.0    &  27.30$\pm$0.27      &      $<$17.0         & 571$\pm$34 & Y & Y & Y & Y &   \\   
L18& LOri-045& 5 35 07.42 \  9 58 22.4& 14.23& 1.33& 2& 12.77& 0.26& M2.0    &  26.89$\pm$0.42      &      $<$17.0         & 554$\pm$23 & Y & Y & Y & Y &   \\   
L19& LOri-046& 5 34 26.08 \  9 51 49.4& 14.36& 1.28& 2& 13.03& 0.23& M2.0    & $-$0.28$\pm$0.43     &                      & 148$\pm$64 & N & N & N & N &   \\    
L20& LOri-048& 5 35 12.56 \  9 53 11.1& 14.41& 1.37& 2& 12.89& 0.27& M2.5    &  27.04$\pm$0.44      &      $<$17.0         & 545$\pm$12 & Y & Y & Y & Y &   \\   
L21& LOri-049& 5 35 01.00 \  9 49 36.1& 14.50& 1.27& 2& 13.17& 0.34& M2.0    &  24.06$\pm$0.36      &                       & 64$\pm$13 & N & N & N & N &   \\   
L22& LOri-050& 5 34 56.40 \  9 55 04.6& 14.54& 1.36& 2& 12.88& 0.29& M2.5    & \multicolumn{1}{c}{?}&                       & 596$\pm$58 & ? & Y & Y & Y?& SB\\    
L23& DM032   & 5 35 18.38 \ 10 02 38.4& 14.69& 1.56& 1& 13.07& 0.26& M1.5    &  26.98$\pm$0.34      &       $<$17.0         & 589$\pm$16 & Y & Y & Y & Y &   \\   
L24& LOri-053& 5 34 36.73 \  9 52 58.3& 14.72& 1.36& 2& 13.17& 0.24& M2.5    &  27.00$\pm$0.42      &     $<$18.3           & 561$\pm$30 & Y & Y & Y & Y &   \\   
L25& LOri-055& 5 35 21.43 \  9 49 56.6& 14.76& 1.36& 2& 13.18& 0.23& M2.5    &  26.70$\pm$0.39      &      $<$17.0          & 602$\pm$33 & Y & Y & Y & Y &   \\   
L26& LOri-056& 5 34 58.37 \  9 53 47.1& 14.87& 1.56& 2& 13.21& 0.30& M3.0    &  26.84$\pm$0.56      &      $<$19.3         & 576$\pm$23 & Y & Y & Y & Y &   \\   
L27& LOri-057& 5 35 11.32 \ 10 00 50.2& 15.04& 1.59& 2& 13.41& 0.28& M3.5    &  26.99$\pm$0.49      &      $<$17.0         & 677$\pm$19 & Y & Y & Y & Y &   \\   
L28& LOri-060& 5 35 20.00 \  9 49 06.2& 15.14& 1.42& 2& 13.60& 0.30& M2.5    &  27.22$\pm$0.45      &      $<$18.5         & 599$\pm$28 & Y & Y & Y & Y &   \\   
L29& LOri-061& 5 35 18.18 \  9 52 24.2& 15.15& 1.43& 2& 13.53& 0.31& M2.5    &  26.80$\pm$0.47      & 17.9$^{+1.4}_{-1.8}$ & 591$\pm$30 & Y & Y & Y & Y &   \\   
L30& LOri-062& 5 35 15.33 \  9 48 37.0& 15.16& 1.46& 2& 13.63& 0.28& M3.0    &  26.80$\pm$0.47      &      $<$17.0         & 648$\pm$48 & Y & Y & Y & Y &   \\   
L31& LOri-068& 5 34 48.02 \  9 43 26.2& 15.20& 1.56& 2& 13.52& 0.27& M3.0    &  29.37$\pm$0.47      &      $<$17.0         & 698$\pm$25 & N & Y & Y & Y?&   \\   
L32& LOri-069& 5 34 43.97 \  9 48 35.6& 15.20& 1.69& 2& 13.38& 0.34& M3.5    & \multicolumn{1}{c}{?}&                      & 516$\pm$57 & ? & Y & Y & Y?& SB\\
L33& LOri-075& 5 34 55.22 \ 10 00 34.7& 15.23& 1.72& 2& 13.40& 0.26& M5.0$^*$&  26.02$\pm$0.82      & 61.3$^{+11.5}_{-4.9}$& 455$\pm$44 & Y & Y & Y & Y &   \\
L34& LOri-063& 5 35 19.14 \  9 54 42.4& 15.34& 1.46& 2& 13.76& 0.41& M3.0    &  27.19$\pm$0.47      &      $<$18.2         & 505$\pm$52 & Y & Y & Y & Y &   \\   
L35& LOri-076& 5 35 10.96 \  9 57 43.8& 15.81& 1.58& 2& 14.22& 0.33& M3.0    &  27.52$\pm$0.50      &      $<$17.0         & 608$\pm$19 & Y & Y & Y & Y &   \\   
L36& LOri-079& 5 34 48.26 \  9 59 53.9& 16.00& 1.51& 2& 14.22& 0.20& M3.0    &  27.03$\pm$0.51      &      $<$17.0         & 607$\pm$59 & Y & Y & Y & Y &   \\   
L37& LOri-080& 5 35 30.05 \  9 59 25.5& 16.01& 1.50& 2& 13.80& 0.31& M3.0    &  25.74$\pm$0.81      &60.2$^{+21.3}_{-5.9}$ & 523$\pm$20 & Y & Y & Y & Y &   \\   
L38& LOri-083& 5 35 43.41 \  9 54 26.8& 16.02& 1.54& 2& 14.27& 0.26& M3.0    &  26.32$\pm$0.57      &19.1$^{+7.9}_{-2.6}$  & 585$\pm$77 & Y & Y & Y & Y &   \\   
L39& LOri-087& 5 34 33.77 \  9 55 34.2& 16.09& 1.45& 2& 14.19& 0.32& M4.5$^*$&  27.25$\pm$0.57      &18.4$^{+6.0}_{-1.7}$  & 609$\pm$42 & Y & Y & Y & Y &   \\
L40& LOri-088& 5 34 49.50 \  9 58 46.8& 16.10& 1.68& 2& 14.14& 0.31& M3.5    &  26.75$\pm$0.53      &      $<$17.0         &  660$\pm$42 & Y & Y & Y & Y &   \\   
L41& LOri-092& 5 35 50.97 \  9 51 03.5& 16.19& 1.65& 2& 14.44& 0.30& M3.5    &  26.66$\pm$0.57      & 19.8$^{+2.8}_{-2.6}$ & 661$\pm$53 & Y & Y & Y & Y &   \\   
L42& LOri-093& 5 34 41.20 \  9 50 16.3& 16.21& 1.61& 2& 14.46& 0.24& M3.5    &  27.00$\pm$0.56      &      $<$18.3         & 623$\pm$61 & Y & Y & Y & Y &   \\   
L43& LOri-094& 5 34 43.17 \ 10 01 59.8& 16.28& 1.75& 2& 14.40& 0.37& M4.0    &  26.36$\pm$0.80      & 54.8$^{+5.5}_{-8.2}$ & 538$\pm$204& Y & Y & Y & Y &   \\   
L44& LOri-095& 5 35 24.18 \  9 55 15.4& 16.35& 1.61& 2& 14.56& 0.30& M6.0$^*$&  27.31$\pm$0.55      &      $<$19.7         & 559$\pm$30 & Y & Y & Y & Y &   \\
L45& LOri-096& 5 35 11.13 \  9 57 19.6& 16.37& 1.65& 2& 14.63& 0.33& M3.5    &  26.92$\pm$0.55      &      $<$19.1         & 553$\pm$48 & Y & Y & Y & Y &   \\   
L46& LOri-100& 5 35 00.10 \  9 46 14.0& 16.43& 1.65& 2& 14.77& 0.22& M3.5    &  28.04$\pm$0.46      &      $<$17.0         & 717$\pm$43 & Y & Y & Y & Y &   \\   
L47& LOri-102& 5 35 22.02 \  9 52 52.3& 16.50& 1.74& 2& 14.63& 0.27& M4.0    &  28.10$\pm$0.53      &      $<$17.0         & 584$\pm$62 & Y & Y & Y & Y &   \\   
L48& LOri-105& 5 34 17.58 \  9 52 29.7& 16.75& 1.83& 2& 14.92& 0.35& M4.0    &  26.45$\pm$0.70      &      $<$20.0         & \multicolumn{1}{c}{Y}& Y & Y & Y & Y & \\   
L49& LOri-106& 5 35 28.77 \  9 54 10.2& 16.76& 1.72& 2& 14.78& 0.42& M4.0    &  26.83$\pm$0.54      &      $<$17.0         & 636$\pm$132& Y & Y & Y & Y &   \\   
\end{longtable}
\end{landscape}
}

\longtab{4}{
 \begin {longtable}{ccccccccc}
\caption{\label{tab:acc_sori} Accretion properties of the $\sigma$ Ori
cluster members}\\
\hline \hline
ID &   H$\alpha$ pEW  & H$\alpha$ 10\%width &
Log(\.{M$_{acc}$}) &NII 6583 \AA & HeI 6678 \AA & SII 6716 \AA& SII 6731 \AA & Disk\\
  &   (\AA)&   (km/s) &(M$_{\odot}$/yr) & (\AA)&  (\AA)& (\AA)  &  (\AA)& class$^{\,a}$\\
\hline 
\endfirsthead
\caption{continued} \\
\hline
ID &   H$\alpha$ pEW  & H$\alpha$ 10\%width &
Log(\.{M$_{acc}$}) &NII 6583 \AA & HeI 6678 \AA & SII 6716 \AA& SII 6731 \AA & Disk\\
  &   (\AA)& (km/s) &(M$_{\odot}$/yr) & (\AA)&  (\AA)& (\AA)  &  (\AA)& class$^{\,a}$\\
\hline 
\endhead
\hline 
\endfoot
\hline
\multicolumn{9}{l}{$a$: disk clasification from \cite{Hernandez2007ApJ}: III=diskless, II=thick disk,}\\
\multicolumn{9}{l}{I= class I candidate, EV=evolved disk}\\
\multicolumn{9}{l}{$b$: splitted in two separate lines}\\
\multicolumn{9}{l}{$c$: classified as a transition disk candidate using the whole SED slope
(see sec. 4.3 in \cite{Hernandez2007ApJ})}\\
\multicolumn{9}{l}{$d$: classified as a class II using the whole SED slope
(see sec. 4.3 in \cite{Hernandez2007ApJ})}\\
\endlastfoot
S02 &   -2.16$\pm$ 0.14 &    146$\pm$12 &              -  &       $\la$0.05 &       $\la$0.05 &             $\la$0.05 &             $\la$0.05 &             III \\
S03 &   -1.39$\pm$ 0.10 &    210$\pm$30 &              -  &       $\la$0.05 &       $\la$0.05 &             $\la$0.05 &             $\la$0.05 &             III \\
S04 &   -6.91$\pm$ 0.14 &            -  &              -  &       $\la$0.10 &       $\la$0.10 &             $\la$0.10 &             $\la$0.10 &             III \\
S05 &  -31.93$\pm$ 1.45 &    500$\pm$20 &   -8.0$\pm$ 0.5 &       $\la$0.05 &       $\la$0.05 &             $\la$0.05 &             $\la$0.05 &              II \\
S06 &  -13.19$\pm$ 1.38 &    504$\pm$57 &   -8.0$\pm$ 0.7 &            0.14 &       $\la$0.03 &             $\la$0.03 &             $\la$0.03 &      EV$^{\,c}$ \\
S07 &   -0.94$\pm$ 0.05 &    159$\pm$20 &              -  &       $\la$0.04 &       $\la$0.04 &             $\la$0.04 &             $\la$0.04 &             III \\
S10 &   -1.89$\pm$ 0.21 &    216$\pm$39 &              -  &       $\la$0.05 &       $\la$0.05 &             $\la$0.05 &             $\la$0.05 &             III \\
S12 &   -5.93$\pm$ 0.14 &    342$\pm$31 &   -9.6$\pm$ 0.5 &            0.39 &       $\la$0.03 &             $\la$0.03 &                  0.04 &              II \\
S13 &  -25.11$\pm$ 0.72 &    539$\pm$20 &   -7.7$\pm$ 0.5 &            0.07 &       $\la$0.06 &             $\la$0.06 &                  0.10 &              II \\
S14 &  -40.03$\pm$ 2.80 &    424$\pm$15 &   -8.8$\pm$ 0.4 &            0.27 &            0.26 &             $\la$0.05 &             $\la$0.05 &              II \\
S15 &  -17.17$\pm$ 0.80 &    226$\pm$ 4 &  -10.7$\pm$ 0.3 &            0.28 &       $\la$0.07 &             $\la$0.07 &             $\la$0.07 &              II \\
S16 &   -1.46$\pm$ 0.17 &    169$\pm$36 &              -  &       $\la$0.03 &       $\la$0.03 &             $\la$0.03 &             $\la$0.03 &             III \\
S17 &  -17.14$\pm$ 1.66 &    214$\pm$17 &  -10.8$\pm$ 0.4 &            0.05 &       $\la$0.03 &                  0.04 &             $\la$0.03 &              II \\
S18 &   -1.25$\pm$ 0.05 &    143$\pm$27 &              -  &       $\la$0.03 &       $\la$0.03 &             $\la$0.03 &             $\la$0.03 &             III \\
S19 &   -2.45$\pm$ 0.18 &    155$\pm$20 &              -  &       $\la$0.03 &       $\la$0.03 &             $\la$0.03 &             $\la$0.03 &             III \\
S21 &  -20.94$\pm$ 1.20 &    317$\pm$ 6 &   -9.8$\pm$ 0.4 &       $\la$0.05 &            0.22 &             $\la$0.05 &             $\la$0.05 &              II \\
S23 &   -0.89$\pm$ 0.13 &    338$\pm$88 &   -9.6$\pm$ 0.9 &            0.05 &       $\la$0.05 &             $\la$0.05 &             $\la$0.05 &              II \\
S24 &   -3.59$\pm$ 0.29 &    178$\pm$27 &              -  &       $\la$0.05 &       $\la$0.05 &             $\la$0.05 &             $\la$0.05 &             III \\
S26 &   -4.65$\pm$ 0.13 &            -  &              -  &       $\la$0.03 &       $\la$0.03 &             $\la$0.03 &             $\la$0.03 &             III \\
S28 &   -5.60$\pm$ 0.72 &    178$\pm$31 &              -  &       $\la$0.10 &       $\la$0.10 &             $\la$0.10 &             $\la$0.10 &             III \\
S29 &                -  &            -  &              -  &              -  &              -  &                    -  &                    -  &              EV \\
S32 &  -25.76$\pm$ 0.79 &    422$\pm$15 &   -8.8$\pm$ 0.4 &            0.66 &            0.35 &     0.06/0.08$^{\,b}$ &     0.19/0.19$^{\,b}$ &       I$^{\,d}$ \\
S33 &  -22.97$\pm$ 0.49 &    445$\pm$23 &   -8.6$\pm$ 0.5 &       $\la$0.08 &            0.11 &             $\la$0.08 &             $\la$0.08 &              II \\
S34 &  -10.19$\pm$ 0.92 &    317$\pm$18 &   -9.8$\pm$ 0.4 &       $\la$0.05 &       $\la$0.05 &             $\la$0.05 &             $\la$0.05 &              II \\
S35 &   -3.78$\pm$ 0.14 &    141$\pm$24 &              -  &       $\la$0.09 &       $\la$0.09 &             $\la$0.09 &             $\la$0.09 &             III \\
S39 &   -5.02$\pm$ 0.30 &    134$\pm$16 &              -  &       $\la$0.10 &       $\la$0.10 &             $\la$0.10 &             $\la$0.10 &             III \\
S41 &  -16.48$\pm$ 0.76 &    191$\pm$10 &        $<$ -11  &            2.34 &            0.19 &                  1.06 &                  1.57 &      EV$^{\,c}$ \\
S43 &  -27.43$\pm$ 2.36 &            -  &              -  &       $\la$0.07 &            0.24 &             $\la$0.07 &             $\la$0.07 &              II \\
S44 & -197.57$\pm$11.64 &    383$\pm$ 4 &   -9.2$\pm$ 0.4 &            1.14 &            0.14 &             $\la$0.08 &             $\la$0.08 &              II \\
S45 &   -6.71$\pm$ 0.23 &    189$\pm$22 &              -  &       $\la$0.10 &            0.11 &             $\la$0.10 &             $\la$0.10 &             III \\
S47 &   -0.73$\pm$ 0.02 &    237$\pm$14 &              -  &       $\la$0.05 &       $\la$0.05 &             $\la$0.05 &             $\la$0.05 &             III \\
S49 &   -4.73$\pm$ 0.34 &    127$\pm$22 &              -  &       $\la$0.08 &       $\la$0.08 &             $\la$0.08 &             $\la$0.08 &             III \\
S51 &   -3.39$\pm$ 0.22 &    157$\pm$12 &        $<$ -11  &       $\la$0.05 &       $\la$0.05 &             $\la$0.05 &             $\la$0.05 &              EV \\
S52 &  -20.86$\pm$ 1.68 &    251$\pm$14 &  -10.5$\pm$ 0.4 &       $\la$0.10 &            0.24 &             $\la$0.10 &             $\la$0.10 &              II \\
S53 &                -  &            -  &              -  &              -  &              -  &                    -  &                    -  &             III \\
S55 &   -1.12$\pm$ 0.08 &     93$\pm$18 &              -  &       $\la$0.03 &       $\la$0.03 &             $\la$0.03 &             $\la$0.03 &             III \\
S56 &   -3.89$\pm$ 0.14 &    132$\pm$21 &              -  &       $\la$0.10 &       $\la$0.10 &             $\la$0.10 &             $\la$0.10 &             III \\
S57 &   -1.40$\pm$ 0.17 &    125$\pm$ 5 &              -  &       $\la$0.09 &       $\la$0.09 &                  0.10 &             $\la$0.09 &             III \\
S58 &   -4.45$\pm$ 0.27 &    116$\pm$ 6 &              -  &       $\la$0.08 &       $\la$0.08 &             $\la$0.08 &             $\la$0.08 &             III \\
S59 &   -2.58$\pm$ 0.04 &     89$\pm$13 &              -  &       $\la$0.10 &       $\la$0.10 &             $\la$0.10 &             $\la$0.10 &             III \\
S60 &   -3.98$\pm$ 0.12 &    121$\pm$11 &              -  &            0.20 &       $\la$0.10 &                  0.36 &                  0.20 &             III \\
S62 &  -10.62$\pm$ 0.57 &    237$\pm$17 &              -  &            0.10 &       $\la$0.03 &                  0.09 &                  0.04 &             III \\
S63 &  -14.11$\pm$ 0.51 &    299$\pm$17 &  -10.0$\pm$ 0.4 &       $\la$0.09 &            0.31 &             $\la$0.09 &             $\la$0.09 &              II \\
S64 &   -4.39$\pm$ 0.10 &    102$\pm$12 &              -  &       $\la$0.08 &       $\la$0.08 &             $\la$0.08 &             $\la$0.08 &             III \\
S65 &   -4.35$\pm$ 0.60 &    223$\pm$42 &  -10.7$\pm$ 0.5 &       $\la$0.08 &       $\la$0.08 &             $\la$0.08 &             $\la$0.08 &              II \\
S67 &   -2.12$\pm$ 0.05 &    114$\pm$19 &              -  &       $\la$0.08 &       $\la$0.08 &             $\la$0.08 &             $\la$0.08 &             III \\
S68 &   -8.79$\pm$ 0.84 &    248$\pm$11 &  -10.5$\pm$ 0.4 &            0.67 &            0.32 &             $\la$0.08 &             $\la$0.08 &              II \\
S69 &  -10.63$\pm$ 0.65 &    379$\pm$14 &   -9.2$\pm$ 0.4 &            0.13 &            0.22 &                  0.25 &                  0.12 &              II \\
S71 & -113.26$\pm$ 6.09 &    379$\pm$ 5 &   -9.2$\pm$ 0.4 &            6.48 &            0.57 &                  0.44 &                  1.24 &              II \\
S72 &   -3.80$\pm$ 0.24 &    203$\pm$23 &  -10.9$\pm$ 0.4 &            0.09 &       $\la$0.05 &                  0.14 &                  0.09 &              II \\
S73 &  -46.98$\pm$ 3.86 &    207$\pm$ 2 &  -10.9$\pm$ 0.3 &            2.05 &            1.13 &             $\la$0.08 &             $\la$0.08 &              II \\
S74 &  -32.72$\pm$ 1.70 &    372$\pm$13 &   -9.3$\pm$ 0.4 &            0.54 &            0.98 &                  0.21 &                  0.55 &              II \\
S75 &   -4.60$\pm$ 0.33 &    102$\pm$ 9 &              -  &       $\la$0.10 &       $\la$0.10 &             $\la$0.10 &             $\la$0.10 &             III \\
S76 &   -4.75$\pm$ 0.29 &    162$\pm$12 &        $<$ -11  &       $\la$0.10 &       $\la$0.10 &             $\la$0.10 &             $\la$0.10 &              II \\
S78 &  -19.65$\pm$ 0.19 &    401$\pm$26 &   -9.0$\pm$ 0.5 &       $\la$0.10 &            0.50 &             $\la$0.10 &             $\la$0.10 &      EV$^{\,c}$ \\
S79 &  -15.87$\pm$ 1.64 &    134$\pm$ 9 &              -  &       $\la$0.08 &            0.21 &                  0.14 &             $\la$0.08 &             III \\
S80 &   -5.76$\pm$ 0.48 &    194$\pm$ 6 &        $<$ -11  &       $\la$0.08 &            0.17 &             $\la$0.08 &             $\la$0.08 &              II \\
S81 &   -8.18$\pm$ 0.61 &    132$\pm$11 &              -  &            0.17 &       $\la$0.10 &                  0.23 &                  0.12 &             III \\
S83 &   -9.71$\pm$ 1.30 &    132$\pm$ 9 &              -  &            0.14 &            0.14 &                  0.20 &                  0.11 &             III \\
S87 &  -36.32$\pm$ 1.54 &    226$\pm$ 4 &  -10.7$\pm$ 0.3 &            4.07 &            0.18 &                  0.41 &                  0.49 &              II \\
S88 &   -7.16$\pm$ 0.52 &    132$\pm$14 &              -  &       $\la$0.10 &       $\la$0.10 &             $\la$0.10 &             $\la$0.10 &             III \\
S92 &  -14.37$\pm$ 0.49 &    157$\pm$ 6 &        $<$ -11  &            0.31 &            0.43 &                  0.65 &                  0.32 &              II \\
S93 & -113.52$\pm$ 6.42 &     70$\pm$ 1 &        $<$ -11  &           49.68 &            0.76 &                  1.47 &                  2.92 &              II \\
S94 &  -82.67$\pm$11.52 &    157$\pm$ 3 &        $<$ -11  &            6.34 &            1.03 &                  0.30 &                  0.62 &              II \\
S98 &  -16.14$\pm$ 0.65 &    333$\pm$ 2 &   -9.7$\pm$ 0.4 &            0.14 &            0.14 &                  0.12 &                  0.06 &              II \\
\end{longtable}
}

\longtab{5}{
 \begin {longtable}{ccccccccc}
\caption{\label{tab:acc_lor} Accretion properties of the $\lambda$
Ori cluster members }\\
\hline \hline
ID &   H$\alpha$ pEW  & H$\alpha$ 10\%width &
Log(\.{M$_{acc}$}) &NII 6583 \AA & HeI 6678 \AA & SII 6716 \AA& SII 6731 \AA & Disk\\
  &   (\AA)& (km/s) &(M$_{\odot}$/yr) & (\AA)&  (\AA)& (\AA)  &  (\AA)& class$^{\,a}$\\
\hline 
\endfirsthead
\caption{continued} \\
\hline
ID &   H$\alpha$ pEW  & H$\alpha$ 10\%width &
Log(\.{M$_{acc}$}) &NII 6583 \AA & HeI 6678 \AA & SII 6716 \AA& SII 6731 \AA & Disk\\
  &   (\AA)& (km/s) &(M$_{\odot}$/yr) & (\AA)&  (\AA)& (\AA)  &  (\AA)& class$^{\,a}$\\
\hline 
\endhead
\hline 
\endfoot
\hline
\multicolumn{9}{l}{$a$: disk classification from \cite{Barrado2007ApJ}: - =Spitzer non-members, III=diskless, }\\
\multicolumn{9}{l}{II=thick disk, EV=evolved disk}\\
\endlastfoot
L01 &  -1.13$\pm$0.15 &            -  &              -  &      $<$0.04 &      $<$0.04 &      $<$0.04 &      $<$0.04 &  III \\
L02 &  -0.50$\pm$0.03 &    137$\pm$15 &        $<$ -11  &      $<$0.03 &      $<$0.03 &      $<$0.03 &      $<$0.03 &    - \\
L03 &  -1.58$\pm$0.13 &            -  &              -  &      $<$0.03 &      $<$0.03 &      $<$0.03 &      $<$0.03 &  III \\
L04 &  -1.00$\pm$0.08 &    155$\pm$13 &        $<$ -11  &      $<$0.03 &      $<$0.03 &      $<$0.03 &      $<$0.03 &    - \\
L06 &  -6.96$\pm$0.69 &    219$\pm$43 &              -  &      $<$0.03 &         0.12 &      $<$0.03 &      $<$0.03 &  III \\
L07 &  -2.00$\pm$0.12 &    137$\pm$20 &              -  &      $<$0.05 &      $<$0.05 &      $<$0.05 &      $<$0.05 &  III \\
L08 &  -5.30$\pm$0.30 &    198$\pm$18 &              -  &      $<$0.03 &      $<$0.03 &      $<$0.03 &      $<$0.03 &  III \\
L09 &  -2.44$\pm$0.23 &    155$\pm$22 &        $<$ -11  &      $<$0.03 &      $<$0.03 &      $<$0.03 &      $<$0.03 &    - \\
L10 &              -  &            -  &              -  &           -  &           -  &           -  &           -  &  III \\
L11 &  -8.47$\pm$0.47 &    215$\pm$25 & -10.80$\pm$0.40 &      $<$0.06 &      $<$0.06 &      $<$0.06 &      $<$0.06 &   II \\
L12 &  -3.32$\pm$0.35 &    146$\pm$34 &              -  &      $<$0.06 &      $<$0.06 &      $<$0.06 &      $<$0.06 &  III \\
L14 &  -3.67$\pm$0.34 &    148$\pm$29 &              -  &      $<$0.06 &      $<$0.06 &      $<$0.06 &      $<$0.06 &  III \\
L15 &  -5.15$\pm$0.62 &    141$\pm$18 &              -  &      $<$0.08 &      $<$0.08 &      $<$0.08 &      $<$0.08 &  III \\
L16 &  -7.05$\pm$0.86 &    306$\pm$52 &              -  &      $<$0.06 &      $<$0.06 &      $<$0.06 &      $<$0.06 &  III \\
L17 &  -1.89$\pm$0.15 &    132$\pm$50 &        $<$ -11  &      $<$0.06 &      $<$0.06 &      $<$0.06 &      $<$0.06 &    - \\
L18 &  -3.04$\pm$0.33 &    123$\pm$27 &              -  &      $<$0.08 &      $<$0.08 &      $<$0.08 &      $<$0.08 &  III \\
L20 &  -3.34$\pm$0.44 &    159$\pm$36 &        $<$ -11  &      $<$0.08 &      $<$0.08 &      $<$0.08 &      $<$0.08 &   EV \\
L22 & -11.14$\pm$0.96 &            -  &              -  &      $<$0.09 &      $<$0.15 &      $<$0.09 &      $<$0.09 &   II \\
L23 &  -4.98$\pm$0.49 &    100$\pm$11 &        $<$ -11  &      $<$0.09 &      $<$0.09 &      $<$0.09 &      $<$0.09 &    - \\
L24 &  -2.93$\pm$0.20 &    130$\pm$22 &              -  &      $<$0.09 &      $<$0.09 &      $<$0.09 &      $<$0.09 &  III \\
L25 &  -6.53$\pm$0.49 &    118$\pm$13 &              -  &      $<$0.09 &      $<$0.09 &      $<$0.09 &      $<$0.09 &  III \\
L26 &  -5.57$\pm$0.81 &    130$\pm$13 &              -  &      $<$0.10 &         0.15 &      $<$0.10 &      $<$0.10 &  III \\
L27 &  -5.20$\pm$0.52 &    107$\pm$11 &              -  &      $<$0.10 &         0.20 &      $<$0.10 &      $<$0.10 &  III \\
L28 &  -4.18$\pm$0.48 &    121$\pm$18 &              -  &      $<$0.08 &      $<$0.08 &      $<$0.08 &      $<$0.08 &  III \\
L29 & -13.97$\pm$1.38 &    274$\pm$15 & -10.20$\pm$0.30 &      $<$0.10 &         0.18 &      $<$0.10 &      $<$0.10 &   II \\
L30 &  -4.41$\pm$0.30 &    123$\pm$22 &        $<$ -11  &      $<$0.10 &      $<$0.10 &      $<$0.10 &      $<$0.10 &   II \\
L31 &  -7.90$\pm$1.26 &    116$\pm$16 &              -  &      $<$0.10 &      $<$0.15 &      $<$0.10 &      $<$0.10 &  III \\
L32 &  -7.39$\pm$1.46 &            -  &              -  &      $<$0.10 &      $<$0.10 &      $<$0.10 &      $<$0.10 &  III \\
L33 & -10.71$\pm$0.98 &    326$\pm$ 9 &              -  &      $<$0.10 &      $<$0.10 &      $<$0.10 &      $<$0.10 &  III \\
L34 & -19.14$\pm$2.53 &    230$\pm$ 4 & -10.70$\pm$0.30 &         3.60 &         0.80 &         0.26 &         0.31 &   II \\
L35 &  -4.92$\pm$0.60 &    121$\pm$20 &              -  &      $<$0.10 &      $<$0.10 &      $<$0.10 &      $<$0.10 &  III \\
L36 &  -5.06$\pm$0.53 &    127$\pm$25 &        $<$ -11  &      $<$0.10 &      $<$0.20 &      $<$0.10 &      $<$0.10 &   EV \\
L37 & -14.47$\pm$1.25 &    287$\pm$41 & -10.10$\pm$0.50 &         0.10 &      $<$0.10 &      $<$0.10 &      $<$0.10 &   EV \\
L38 &  -5.32$\pm$0.68 &    134$\pm$16 &              -  &      $<$0.10 &         0.10 &      $<$0.10 &      $<$0.10 &  III \\
L39 &  -5.56$\pm$0.70 &    125$\pm$18 &        $<$ -11  &      $<$0.10 &      $<$0.20 &      $<$0.10 &      $<$0.10 &   EV \\
L40 &  -9.44$\pm$1.12 &    107$\pm$13 &              -  &      $<$0.10 &         0.17 &      $<$0.10 &      $<$0.10 &  III \\
L41 &  -2.90$\pm$0.34 &    107$\pm$29 &              -  &         0.14 &      $<$0.10 &      $<$0.10 &      $<$0.10 &  III \\
L42 &  -6.25$\pm$0.94 &    137$\pm$29 &              -  &      $<$0.10 &         0.15 &      $<$0.10 &      $<$0.10 &  III \\
L43 & -15.53$\pm$2.23 &    228$\pm$18 &              -  &      $<$0.20 &      $<$0.20 &      $<$0.20 &      $<$0.20 &  III \\
L44 &  -4.96$\pm$0.62 &    105$\pm$ 9 &              -  &      $<$0.25 &      $<$0.25 &      $<$0.25 &      $<$0.25 &  III \\
L45 &  -6.97$\pm$1.10 &    114$\pm$20 &        $<$ -11  &         0.22 &      $<$0.20 &      $<$0.20 &      $<$0.20 &   II \\
L46 &  -8.18$\pm$1.40 &     98$\pm$ 6 &              -  &         0.60 &      $<$0.25 &         0.59 &      $<$0.25 &  III \\
L47 &  -5.94$\pm$0.96 &     93$\pm$ 4 &              -  &         0.25 &      $<$0.10 &         0.25 &         0.16 &  III \\
L48 &              -  &            -  &              -  &           -  &           -  &           -  &           -  &  III \\
L49 & -26.16$\pm$2.40 &    107$\pm$ 4 &        $<$ -11  &         0.70 &         0.40 &         0.40 &         0.30 &   II \\
\end{longtable}
}

\end{document}